\documentclass[reprint,twocolumn,aps,preprintnumbers,superscriptaddress]{revtex4-2}

\usepackage{graphicx}
\usepackage{dcolumn}
\usepackage{bm}
\usepackage{amsthm,amsmath,amssymb}
\usepackage{mathrsfs}
\usepackage{appendix}
\usepackage{amstext}
\usepackage{url}
\usepackage{float}
\usepackage[usenames,dvipsnames]{color}
\usepackage[hidelinks]{hyperref}
\usepackage{booktabs}
\usepackage{algorithm2e}
\usepackage{siunitx}
\usepackage{makecell}
\usepackage{subfigure} 
\usepackage{pifont} 
\usepackage{tabularx} 

\newcommand{\etal}{{\it{et al.~}}}

\newcommand{\be}{\begin{equation}}
\newcommand{\ee}{\end{equation}}

\newcommand{\sket}[1]{{\ensuremath{\lvert#1\rangle}}}
\newcommand{\lket}[1]{{\ensuremath{\left\lvert#1\right\rangle}}}

\newcommand{\ket}[1]{\if@display\lket{#1}\else\sket{#1}\fi}

\newcommand{\sbra}[1]{{\ensuremath{\langle#1\rvert}}}
\newcommand{\lbra}[1]{{\ensuremath{\left\langle#1\right\rvert}}}
\newcommand{\bra}[1]{\if@display\lbra{#1}\else\sbra{#1}\fi}

\newcommand{\sketbra}[2]{{\ensuremath{\lvert #1\rangle\!\langle #2\rvert}d}}
\newcommand{\lketbra}[2]{{\ensuremath{\left\lvert #1\right\rangle\!\!\left\langle #2\right\rvert}}}
\newcommand{\ketbra}[2]{\if@display\lketbra{#1}{#2}\else\sketbra{#1}{#2}\fi}

\theoremstyle{plain}

\theoremstyle{definition}

\hypersetup{colorlinks=true, linkcolor=red, citecolor=blue, urlcolor=black}
 
\begin{document}

\title{Repeater-like asynchronous measurement-device-independent quantum conference key agreement}
\author{Yu-Shuo Lu}\thanks{These authors contributed equally.}
\affiliation{National Laboratory of Solid State Microstructures and School of Physics, Collaborative Innovation Center of Advanced Microstrucstures, Nanjing University, Nanjing 210093, China}
\affiliation{School of Physics and Beijing Key Laboratory of Opto-electronic Functional Materials and Micro-nano Devices, Key Laboratory of Quantum State Construction and Manipulation (Ministry of Education), Renmin University of China, Beijing 100872, China}
\author{Hua-Lei Yin}\email{hlyin@ruc.edu.cn}\thanks{These authors contributed equally.}
\affiliation{School of Physics and Beijing Key Laboratory of Opto-electronic Functional Materials and Micro-nano Devices, Key Laboratory of Quantum State Construction and Manipulation (Ministry of Education), Renmin University of China, Beijing 100872, China}
\affiliation{National Laboratory of Solid State Microstructures and School of Physics, Collaborative Innovation Center of Advanced Microstrucstures, Nanjing University, Nanjing 210093, China}
\affiliation{Beijing Academy of Quantum Information Sciences, Beijing 100193, China}
\affiliation{Yunnan Key Laboratory for Quantum Information, Yunnan University, Kunming 650091, China}
\author{Yuan-Mei Xie}\thanks{These authors contributed equally.}
\affiliation{National Laboratory of Solid State Microstructures and School of Physics, Collaborative Innovation Center of Advanced Microstrucstures, Nanjing University, Nanjing 210093, China}
\affiliation{School of Physics and Beijing Key Laboratory of Opto-electronic Functional Materials and Micro-nano Devices, Key Laboratory of Quantum State Construction and Manipulation (Ministry of Education), Renmin University of China, Beijing 100872, China}
\author{Yao Fu}
\affiliation{Beijing National Laboratory for Condensed Matter Physics and Institute of Physics, Chinese Academy of Sciences, Beijing 100190, China}
\author{Zeng-Bing Chen}\email{zbchen@nju.edu.cn}
\affiliation{National Laboratory of Solid State Microstructures and School of Physics, Collaborative Innovation Center of Advanced Microstrucstures, Nanjing University, Nanjing 210093, China}

\begin{abstract}
Quantum conference key agreement enables secure communication among multiple parties by leveraging multipartite entanglement, which is expected to play a crucial role in future quantum networks. However, its practical implementation has been severely limited by the experimental complexity and low efficiency associated with the requirement for synchronous detection of multipartite entangled states. In this work, we propose a measurement-device-independent quantum conference key agreement protocol that employs asynchronous Greenberger-Horne-Zeilinger state measurement. Our protocol enables a linear scaling of the conference key rate among multiple parties, demonstrating performance comparable to that of the single-repeater scheme in quantum networks. Additionally, we achieve intercity transmission distances with composable security under finite-key conditions. 
By adopting the generalized asynchronous pairing strategy, our approach eliminates the need for complex global phase locking techniques. 
Furthermore, by integrating asynchronous pairing with ring-interference network structure, our method provides insights for various quantum tasks beyond quantum communication, including multiparty computing and quantum repeaters.
\\
\\
\textbf{Keywords}: multiphoton interference, conference key agreement, generalized asynchronous pairing, GHZ entanglement, quantum network 
\end{abstract}

\maketitle
 
\section{Introduction}

Rapidly developing quantum networks are revolutionizing various quantum information processing tasks for multiple parties~\cite{kimble2008quantum,Wehner2018} 
including quantum secure communication {~\cite{bennett2014quantum,dynes2019cambridge,chen2021implementation,cao2024experimental,weng2023beating,lo2012measurement, braunstein2012side,long2002theoretically,zhou2020device,pan2024evolution}}, distributed quantum computing~\cite{van2016path}, and distributed quantum sensing~\cite{komar2014quantum,guo2020distributed,zhao2021field}.
Quantum conference key agreement (QCKA) is a promising application of quantum networks, enabling a group of users within a quantum network to efficiently distribute information-theoretically secure conference keys~\cite{bose1998multiparticle,chen2007multi,fu2015long,proietti2021experimental}.

The foundation of QCKA lies in establishing multipartite entanglement~\cite{pan2012multiphoton} among distant participants. Intuitively, this can be achieved by directly distributing the Greenberger-Horne-Zeilinger (GHZ) entangled state~\cite{chen2007multi}. This approach has been extensively studied in various scenarios~\cite{epping2017multi,ribeiro2018fully,cao2021coherent,kaur2020fundamental,holz2020genuine,li2021finite,horodecki2022fundamental,grasselli2018finite,bai2022post,pickston2023conference,liu2023experimental,philip2023multipartite}. However, the practical application of this approach is hindered by the complexity of preparing and distributing high-fidelity entangled states in real quantum networks. A solution to this challenge is measurement-device-independent QCKA (MDI-QCKA)~\cite{fu2015long} and its continuous variable version~\cite{ottaviani2019modular}, which introduce an untrusted relay node architecture to distribute post-selected entanglement states. This architecture resembles a single-node quantum repeater but does not require quantum memory. By leveraging this architecture, one can be immune to all detector attacks and avoid the need for preparing and distributing entangled states. 
Recent experiments have demonstrated three-user MDI-QCKA networks, achieving a key rate of 45.5 bits/s over a 60-km-long fiber link~\cite{yang2024experimental}
and 0.097 bits/s with a channel loss of  21.5 dB loss~\cite{du2024experimental}. 
These experiments confirm the feasibility of using post-selected GHZ states for achieving quantum conferencing over long distances.

However, both the directly distribution of GHZ states and MDI-QCKA face significant challenges, as their key rates decrease exponentially with the number of participants $N$. This is because they rely on $N$-fold coincidences to successfully distribute (post-selected) GHZ states. As a result, their key rates scale as $O(\eta^N)$, where $\eta$ is the channel transmittance between the user and the untrusted node in the quantum network. 
Fundamentally, there is a performance limit on the distribution of multipartite entanglement in repeaterless quantum networks, which establishes an upper bound on the rates at which conference keys can be generated~\cite{das2021universal,pirandola2020general}. 
Specifically, the single-message multicast bound in a star network is given by $R\le-\log_2(1-\eta^2)$~\cite{pirandola2020general}, indicating that the conference key rate cannot surpass the Pirandola-Laurenza-Ottaviani-Banchi (PLOB) bound~\cite{pirandola2017fundamental}. To break the PLOB bound, single-photon interference~\cite{Lucamarini2018} and asynchronous two-photon interference~\cite{xie2022breaking,zeng2022mode} provide feasible solutions for the two-user scenario. 
In the multiparty case, several efforts~\cite{zhao2020phase,grasselli2019conference,Carrara2023overcoming,xie2024multi,li2024asynchronous,li2023breaking} have been made to break the PLOB bound, requiring complex techniques such as simultaneously global phase-locking all users~
{
\cite{zhao2020phase,grasselli2019conference,Carrara2023overcoming,xie2024multi}.} A comparison of recent MDI-QCKA protocols is shown in Table~\ref{tab_compare}.

\begin{table}[b] 
\centering
\caption{
Summary of the key features of recent MDI-QCKA protocols. The symbol \ding{51} (\ding{53}) represents a positive (negative) result. The comparison of the key rate scaling regarding the channel transmittance $\eta$ from one user to the measurement node and the user number $N$ is also provided.
}\label{tab_compare}
\begin{tabular}[b]{l@{\hspace{-8pt}}cccl}
\hline
\hline
            \makecell[l]{~\\~} &
		\makecell[c]{Phase-locking \\independence }&
            \makecell[c]{Finite\\key }&
            \makecell[c]{Scalability\\ to $N\ge4$ }&\makecell[c]{Scaling} \\
			\hline
   	    This work & \ding{51}  & \ding{51} & \ding{51} &          $O(\eta)$ 
            \\
            Fu \etal \cite{fu2015long} & \ding{51} & \ding{53} & \ding{51} &  $O(\eta^N)$  \\
            Zhao \etal \cite{zhao2020phase} & \ding{53} & \ding{53} & \ding{51} & $O(\eta^{N-1})$  \\
            Ottaviani \etal \cite{ottaviani2019modular} & \ding{53} & \ding{51} & \ding{51} & $O(\eta^N)$ \\
             Carrara \etal \cite{Carrara2023overcoming} & \ding{53} & \ding{53} & \ding{51} & $O(\eta)$  \\
            Xie \etal \cite{xie2024multi} & \ding{53} & \ding{51} & \ding{51} & $O(\eta)$  \\
        \hline
		\hline
		\end{tabular}
\end{table}

\begin{figure}[t] 
\includegraphics[width=0.49\textwidth]{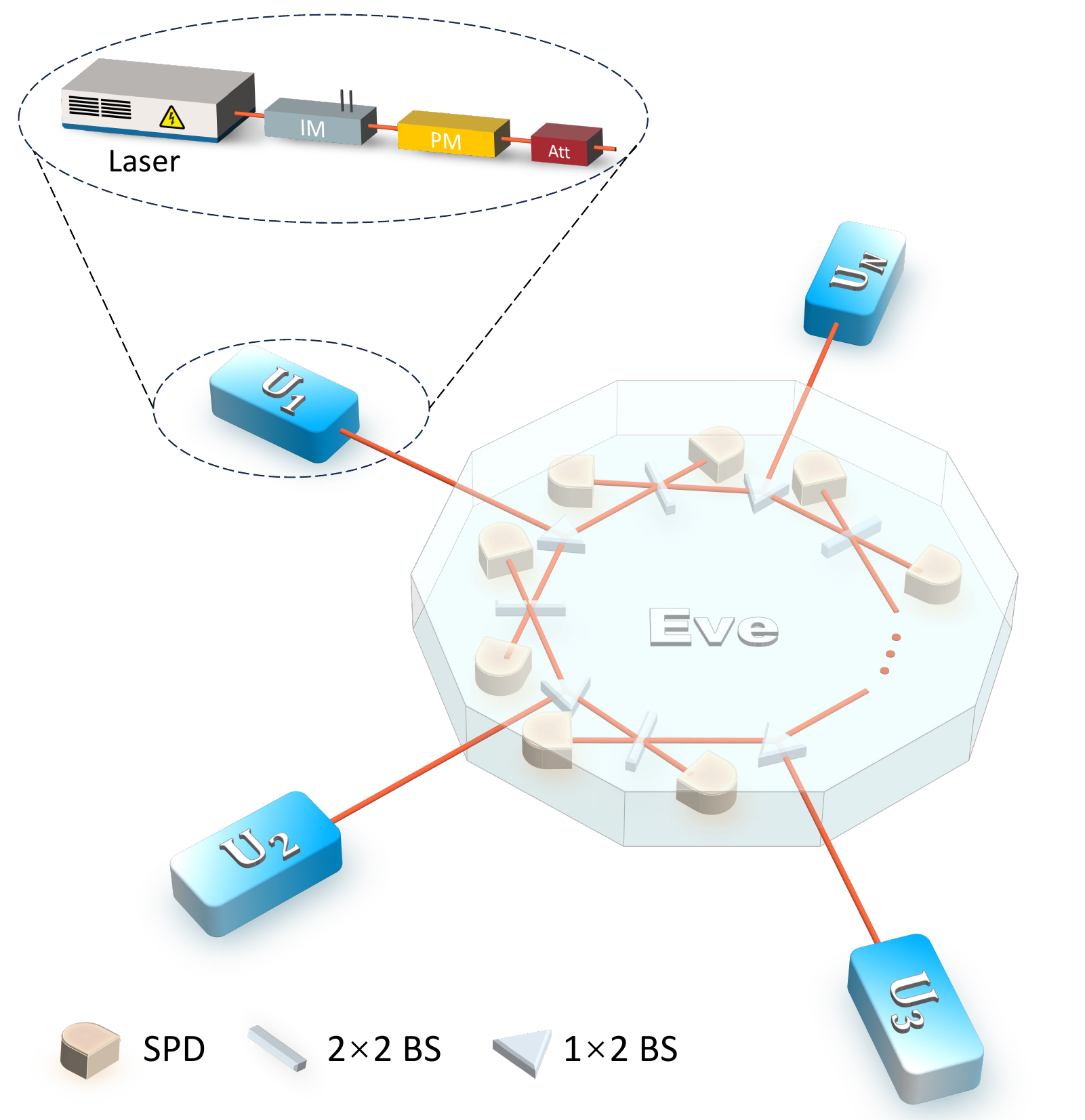}
\caption{Conceptual schematic of the AMDI-QCKA protocol. There are $N$ users connected to an untrusted measurement node, Eve, via optical channels. Each user employs a laser, intensity modulator (IM), phase modulator (PM), and attenuator (Att) to prepare phase-randomized weak coherent pulses. Eve contains $N$ detection ports, collectively forming a $2N$-sided polygonal structure. Each detection port includes a $2\times2$
beam splitter (BS) and two single-photon detectors (SPDs) which are connected to the left and right outputs of BS. The optical pulses from each user are split by the $1\times2$ BS into two parts, which are directed to two adjacent measurement ports, where they interfere with pulses from adjacent users.
}\label{fig1_qcka_diag}
\end{figure}

In this work, we propose a scalable asynchronous MDI-QCKA (AMDI-QCKA) protocol by integrating the generalized asynchronous pairing strategy with the ring-interference network. Our protocol leverages asynchronous multi-photon interference within an untrusted relay node architecture. The implementation of the $N$-user QCKA network involves a $2N$-shaped ring multi-interference network with $N$ detection ports. By pairing $N$ interference events at different ports and different time bins within the coherent time, effective GHZ-measurement events are obtained. This generalized asynchronous pairing strategy removes the need for complex global phase-locking and phase-tracking techniques. By employing phase-randomized weak coherent sources and the decoy-state method~\cite{hwang2003quantum,wang2005beating,lo2005decoy}, we provide a solution for obtaining a secure conference key rate in the finite-size regime with composable security. Our protocol requires only a fixed number of decoy-state intensities, in contrast to {Ref.~\cite{zhao2020phase}} where the number scales with the number of users. 
With sufficient click events within the coherent time, the key rate scales as $O(\eta)$ in the high count rate limit.
Therefore, our protocol can surpass the PLOB bound with simple technical requirements. 



\begin{figure*} 
\includegraphics[width=0.99\textwidth]{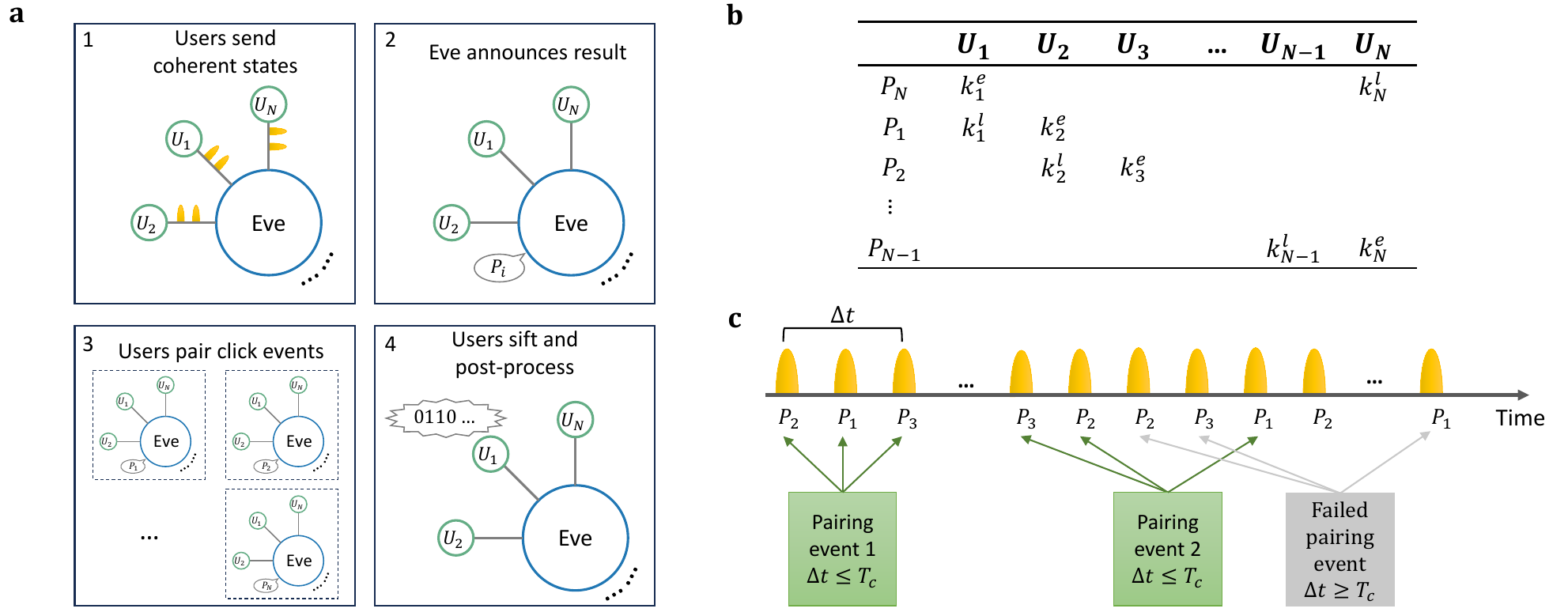}
\caption{{(a) Flowchart of the AMDI-QKCA protocol. All Users send coherent states to a measurement node, Eve, who performs interference measurements and announces the results. Users pair click events, perform sifting and post-processing, and generate secret conference keys.
}
(b) The relationship between the users and the detection ports where interference occurs. Users $U_i$ and $U_{i+1}$ interfere at $P_i$ ($U_N$ and $U_{1}$ interfere at $P_N$), and they register an early and a late time bin, respectively, while the remaining $N-2$ users are not associated with $P_i$.
(c) Schematic diagram of the pairing scheme for $N=3$. A paired event consists of $3$ detection events occurring at $3$ different ports. Starting from the first click event at $P_2$, three consecutive click events occur at different ports, forming the first pairing event. For the second pairing event, repeated click events at the ports $P_2$ and $P_3$ are skipped. If pairing time interval $\Delta t$ between different ports exceed the 
$T_c$, a pairing failure occurs. 
} \label{fig_pairscheme}
\end{figure*}

\section{Protocol description} \label{sec_protocol}
The topology of the $N$-user AMDI-QCKA network is shown in Fig.~\ref{fig1_qcka_diag}. The network consists of $N$ users, labeled $U_1$, $U_2$, $\dots$, $U_N$, and an untrusted intermediate measurement node controlled by Eve. Each user is equipped with a laser, an intensity modulator, a phase modulator and an attenuator, all used to generate phase-randomized weak coherent pulses at the single-photon level. The measurement node, controlled by Eve, can be any structure, but is expected to have $N$ detection ports $P_1$, $P_2$, $\dots$, $P_N$ arranged in a $2N$-sided polygonal configuration. 
Each port $P_i$ ($i\in \{1, 2,..., N\}$) contains a 50/50 $2\times 2$ beam splitter (BS) and two single photon detectors, $L_i$ and $R_i$. The detailed AMDI-QCKA protocol is given as follows, and its flow chart is shown in Fig.~\ref{fig_pairscheme}(a).
 
\textbf{Step 1.} (Preparation): 
For each time bin, each user $U_i$ prepares a phase-randomized weak coherent pulse $\ket{e^{\textbf{i}\theta_i}\sqrt{k_i}}$ with intensity $k_i$, probability $p_{k_{i}}$, random phase $\theta_i=2\pi M_{i}/M$,
with $k_i$ $\in$ $\{\mu_i, \nu_i, o\}$ (representing signal, decoy, and vacuum state, with $\mu_i>\nu_i>o$), where $M_{i} \in \{0,1,...,M-1\}$ denotes the random phase slice. All users then send their pulses to Eve.  

\textbf{Step 2.} (Measurement): Eve performs interference measurement on the received pulses, where each user's pulse interferes with the pulses from its two neighboring users. The pulse from each user is first split by a $1\times2$ BS, and then directed to two neighboring ports. At detection port $P_i$ ($i \neq N$), the pulses from $U_i$ interfere with those from $U_{i+1}$, and at $P_N$, the pulses from $U_N$ interfere with those from $U_1$, as shown in Fig.~\ref{fig_pairscheme}(b). In each time bin, successful click events are recorded only when exactly one detector registers a click. Eve publishes all successful click events and announces which detectors clicked.  The above steps are repeated for a sufficiently large number of rounds.

\textbf{Step 3.} (Pairing): The users pair $N$ successful click events occurring in $N$ different detection ports with a time interval less than the maximum pairing time $T_c$ to form a pairing event, as shown in Fig.~\ref{fig_pairscheme}(c). An example of pairing algorithm is provided in Appendix~A.
At the time bin when $P_i$ ($i \neq N$) records a single click, $U_i$ ($U_{i+1}$) registers a late (early) time bin, denoted by superscript $l(e)$. When $P_N$ clicks, $U_N$ ($U_1$) registers a late (early) time bin. Each user $U_i$ calculates $k_i^{\rm tot} = k_i^{e}+k_i^{l}$, representing the total intensity used in the two time bins when $U_i$'s two neighboring detection ports have single clicks and the global phase difference $\theta_{i}^d = \theta_i^l-\theta_i^e$.  

\begin{table}[b]
\centering
\caption{The key mapping rule in the $\boldsymbol{X}$-basis. $r_i$= 0 (1) represents the detector $L_i$ $(R_i)$ clicks.
} \label{tab_keysift}
\begin{tabular}{cccc}
			\hline
			{$\theta_{g}^d/\pi\mod 2$}& $U_1$'s bit & {$U_i$'s bit  ($i\in\{2,...,N\}$)} \\
			\hline
   	    $ 0$ & $ r_1  \oplus ... \oplus  r_N\oplus0$ & $0$ &\\ 
			$1$&$ r_1   \oplus ... \oplus  r_N\oplus1$& $ 0 $\\
   	 Others  &            {Discard}  & Discard \\
\hline
\end{tabular}
\end{table}

\textbf{Step 4.}  (Sifting): For each pairing event, each user $U_i$ publicly announces $k_i^{\rm tot}$ and $\theta_{i}^d$. The paired event can be represented as $[k_1^{\rm tot},k_2^{\rm tot}, \dots, k_N^{\rm tot}]$. The participants assign $[\mu_1, \mu_2, ..., \mu_N]$ events to $\boldsymbol{Z}$-basis. For pairing events $[2\nu_1, 2\nu_2, ...,2\nu_N]$, they calculate the sum of the global phase difference $\theta_{g}^d =\sum\limits_{i=1}^N \theta_{i}^d$. If $(\theta_{g}^d/\pi) \mod 2 \in \{0,1\}$, the event is assigned to the $\boldsymbol{X}$-basis. For a pairing event in the $\boldsymbol{Z}$-basis, each user $U_i$ obtains a bit value 0 if $k_i^e = \mu_i$ and $k_i^l = o$. Otherwise, a bit value of 1 is assigned.
To estimate the $\boldsymbol{X}$-basis error rate, bit values are determined based on $N$-fold coincidence detection~\cite{greenberger1990bell} and the value of $\theta_{g}^d/\pi \mod 2$, as shown in Table~\ref{tab_keysift}. 

In addition to the above simplified version of key mapping, we also provide a detailed version. Each user decomposes the global random phase $ \theta_i^d $ into $ \vartheta_i + \kappa_i \pi $, where $ \vartheta_i $ represents the reference frame phase and $ \kappa_i \in \{0,1\} $ is the encoding bit. Pairing events that differ in reference frame phase by $ \pi $ or are identical are grouped into a set $ [\vartheta_1, \vartheta_2, \dots, \vartheta_N]$. Considering  $ U_1 $ as the reference, each user determines their own key, which is the value of $ \kappa_i $ for $ i \neq 1 $, and $ U_1 $ computes the secret key by calculating $ \kappa_1 \oplus r_1 \oplus r_2 \oplus \dots \oplus r_N \oplus (\vartheta_g/\pi\mod2 ) $.

\begin{figure*}[t]
\centering
\subfigure{\label{fig_asyNa}
\includegraphics[width=0.45\textwidth]{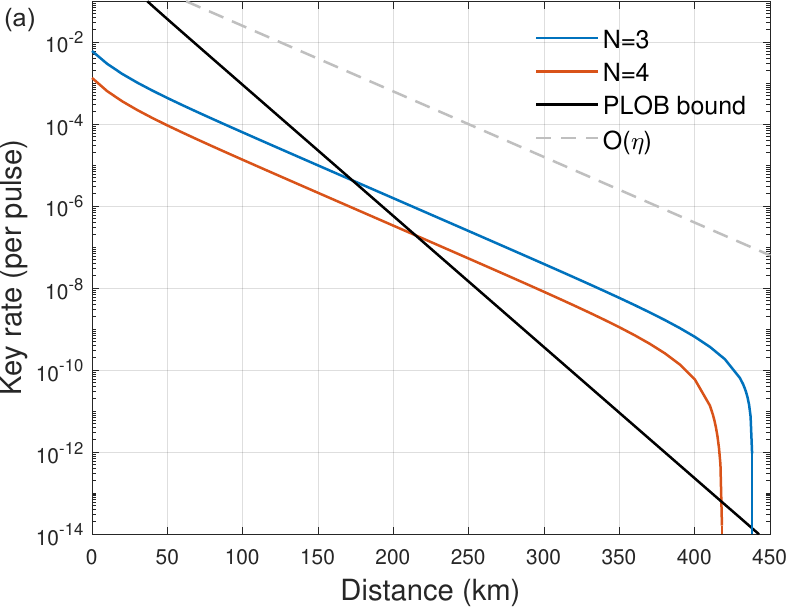}}
\subfigure{\label{fig_asyNb}
\includegraphics[width=0.48\textwidth]{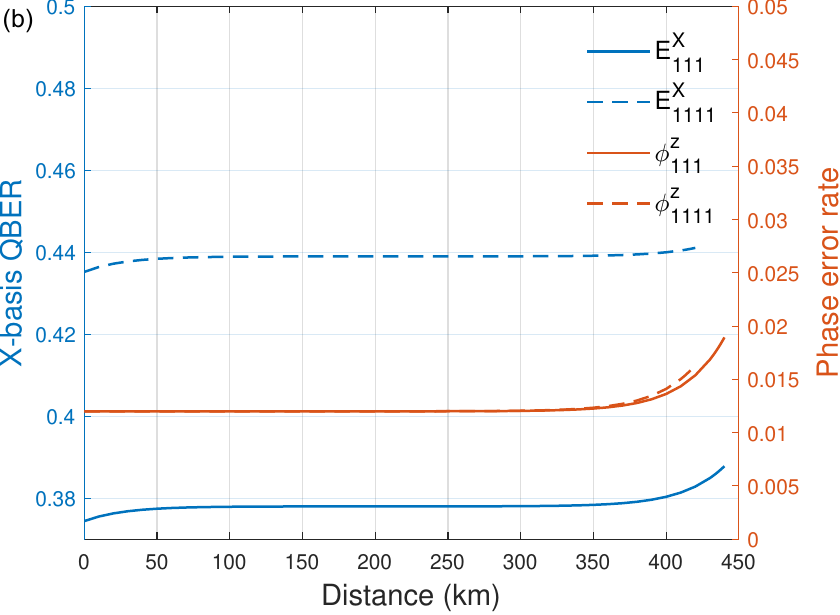}}
\caption{Simulation results for the conference key rate and the error rate are presented as functions of the transmission distance between users and Eve. The three-user and four-user cases in the asymptotic limit with an infinite number of decoy states in symmetric channels are considered. For all simulations, we use the same parameters: fiber channel loss coefficient of 0.16 dB km$^{-1}$, detection efficiency of 85$\%$, dark counting rate $10^{-10}$, $\boldsymbol{X}$-basis misalignment rate $e_d = 1.2\%$, error correction efficiency $f$=1.02. Here, we assume global phase-locking is applied. 
(a) Conference key rate as a function of distance. (b)  $\boldsymbol{Z}$-basis phase error rate $ \phi_{\{1\}^{N}}^z$ and $\boldsymbol{X}$-basis quantum bit error rate (QBER) as functions of distance{
, using dual vertical axes: X-basis QBER on the left, phase error rate on the right}.   
}
\end{figure*}     

\textbf{Step 5.} (Parameter estimation and postprocessing): By using the decoy state method, in the $\boldsymbol{Z}$-basis, they estimate the number of single-photon components
$s_{\{1\}^{N}}^z$ and the phase error rate $\phi_{\{1\}^{N}}^z$. By applying multiparty entanglement-distillation techniques~\cite{chen2007multi,fu2015long}, the final key length in the asymptotic limit is given by
\begin{equation}
\begin{aligned}
l_{\rm asy} = &  s_{\{1\}^{N}}^z\left[1-H_2(\phi_{\{1\}^{N}}^z)\right]- {\rm leak}_{\rm EC},
\end{aligned}
\end{equation} 
where ${\rm leak}_{\rm EC} =\max \limits_{2\le i \le N}[H_2(E^z_{1,i})]fn^z$, $n^z$ is the number of $\boldsymbol{Z}$-basis events, $H_2(x)=-x\log_2x-(1-x)\log_2(1-x)$ is the binary Shannon entropy function, and $f$ is the error correction efficiency. We consider $U_1$'s key as the reference key and $E^z_{1,i}$ is the marginal bit error rate between $U_1$ and $U_i$.

Specifically, the key length formula in the finite-size regime against coherent attacks with composable security $\varepsilon_{\rm cor}$-correctness and $\varepsilon_{\rm sec}$-secrecy when $N =3$ can be written as
\begin{equation}
\begin{aligned}\label{eq_3key_main}
l \ge& ~\underline{s}_{0}^z +\underline{s}_{111}^z[1-H_2(\overline{\phi}_{111}^z)] - {\rm leak}_{\rm EC}
\\
&-\log_2 \frac{4}{\varepsilon_{\rm cor}}  -2\log_2\frac{2}{{
\varepsilon'\hat{\varepsilon}}}
-2\log_2\frac{1}{2\varepsilon_{\rm PA}},
\end{aligned}
\end{equation}
where $\underline{x}$ and $\overline{x}$ are the lower and upper bounds of the observed value $x$, respectively. $\underline{s}_{0}^z$, $\underline{s}_{111}^z$, and $\overline{\phi}_{111}^z$ 
are the number of vacuum events in the $\boldsymbol{Z}$-basis, the number of single-photon components in the $\boldsymbol{Z}$-basis and the corresponding phase error rate, respectively,
$n^z$ is the number of $\boldsymbol{Z}$-basis event. The overall security $\varepsilon_{\rm tot}$ is defined as: $\varepsilon_{\rm tot} = \varepsilon_{\rm sec}+ \varepsilon_{\rm cor}$,  $\varepsilon_{\rm sec} = 2(\varepsilon'+2\varepsilon_e+\hat{\varepsilon})+\varepsilon_0+\varepsilon_3 + \varepsilon_\beta + \varepsilon_{\rm PA}$, where $\varepsilon'$, $\varepsilon_e$ and $\hat{\varepsilon}$ are security parameters, $\varepsilon_0$, $\varepsilon_3$ and $\varepsilon_\beta$ quantify the failure probabilities in estimating the terms of $\underline{s}_0^z$, $\underline{s}_{111}^z$, and $\overline{\phi}_{111}^z$, respectively.
Detailed formulas for composable security and decoy-state estimation are provided in Appendix~C and~D, respectively. 
 
\section{Performance} \label{sec_performance}
We first provide an intuitive understanding of the scaling of the AMDI-QCKA protocol.  We define the key rate as $R:={l}/{\mathcal{N}}$, where $\mathcal{N}$ is the total number of pulses transmitted by each user. 
The raw key rate mainly depends on the number of pairing events $n_{\rm tot}$. 
We analyze $n_{\rm tot}$ in the high count rate regime where a sufficient number of click events are registered at each detection port within the coherence time. 
In each time bin, each user sends a weak coherent pulse to the intermediate node, with channel transmittance $\eta$. 
The probability of a single click event is given by $q_{\rm tot} \approx 1 - (1-\mu_{\rm mean}\eta)^N\approx
N \mu_{\rm mean}\eta$, where $\mu_{\rm mean}$ is the mean intensity. During the pairing step, $N$ single click events are paired, introducing a factor of $1/N$, and 
the number of pairing events is $n_{\rm tot} = \mathcal{N}q_{\rm tot}/N \approx \mathcal{N} \mu_{\rm mean}\eta$. Therefore, the key rate exhibits a repeater-like scaling $O(\eta)$. A detailed analysis is provided in Appendix~F.

The numerical simulation of the conference key rate of the AMDI-QCKA protocol in symmetric channels is presented below. Detailed simulation formulas are provided in Appendix~E. 
Users select intensities from the same set $[\mu, \nu, o]$.
Intensities and the corresponding probabilities are globally optimized. To demonstrate the performance in the ideal case, we first consider the asymptotic key rate with an infinite number of decoy states, and we assume that phase-locking is employed. The use of phase-locking implies that the coherence time between users becomes infinite, resulting in an infinite $T_c$. Therefore, all detection events are paired. As shown in Fig.~\ref{fig_asyNa}, in both the $N=3$ and $N=4$ cases, our protocol can surpass the PLOB bound and exhibits linear scaling of key rate as $\eta$ up to 400 km.  
The phase error rate and the $\boldsymbol{X}$-basis quantum bit error rate (QBER) are shown in Fig.~\ref{fig_asyNb}. An intrinsic $\boldsymbol{X}$-basis QBER of about $37.5\%$ for $N=3$ and $43\%$ for $N=4$ arises from multi-photon components in the coherent state. In a paired event, the pulses transmitted by user $U_i$ in two time bins constitute a part of the joint GHZ measurement event. If at least one user emits a zero-photon state over two time bins and at least one user emits multiple photons, an erroneous GHZ state measurement event occurs. 
Regarding the phase error rate, there is no intrinsic error, since the single-photon components, where each of the $N$ users emits a single photon, can be accurately estimated.
The result is consistent with that of MDI-QCKA~\cite{fu2015long}. 
 
\begin{figure}[t]
\centering
\includegraphics[width=0.99\linewidth]{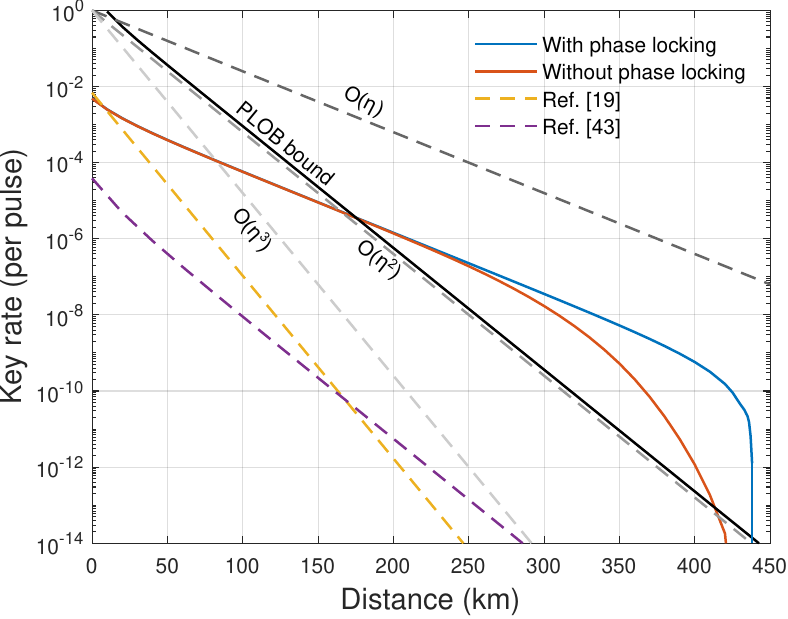}
	\caption{Conference key rate of the AMDI-QCKA protocol in asymptotic limit with three decoy states and three users. We simulate our protocol with and without global phase locking, and  compare this work with Ref.\cite{fu2015long} and \cite{zhao2020phase}. Before 300 km, the key rate of our protocol is parallel with the line of $O(\eta)$, and Ref.\cite{fu2015long} and \cite{zhao2020phase} are parallel with $O(\eta^3)$ and $O(\eta^2)$, respectively. The system frequency is $4\times10^9$ Hz.
    The maximum frequency difference between users' lasers is $10$ Hz, and fiber phase drift rate is $3000$ rad/s, corresponding to $T_c\approx 300~\mu s$. 
}\label{fig_SKR_comp} 
\end{figure}
We also compared the key rate of AMDI-QCKA with that of MDI-QCKA~\cite{fu2015long} and phase-matching 
QCKA~\cite{zhao2020phase}, as shown in Fig.~\ref{fig_SKR_comp}.
We consider a scenario with three users employing three decoy states in the asymptotic regime. For our protocol, two cases are considered: with and without global phase-locking. When the phase-locking is removed, the number of pairing events  becomes a function of distance and $T_c$, where $T_c$ is globally optimized in simulation. 
MDI-QCKA requires synchronous $3$-multiple coincidences for a successful GHZ measurement event, resulting in a key rate scaling of $O(\eta^3)$. Phase-matching QCKA requires the coincidence clicks of $2$ measurement branches, leading to a key rate scaling of $O(\eta^{2})$. While global phase-locking is necessary to phase-matching QCKA, but is not required by MDI-QCKA. With the linear scaling of the AMDI-QCKA, our protocol surpasses the PLOB bound and achieves a maximum transmission distance of over 400 km. The key rates, regardless of whether  phase-locking is applied, are almost the same for distances $L<250$ km, as sufficient detection events can be found within $T_c$. At 300 km,  our protocol exhibits a key rate advantage of approximately six orders of magnitude. Additionally, if we set a cutoff threshold of $10^{-10}$ for real-life consideration, our protocol achieves a transmission distance exceeding 400 km, whereas the other two protocols remain below 170 km. 

\begin{figure}[t]
\centering
\includegraphics[width=1\linewidth]{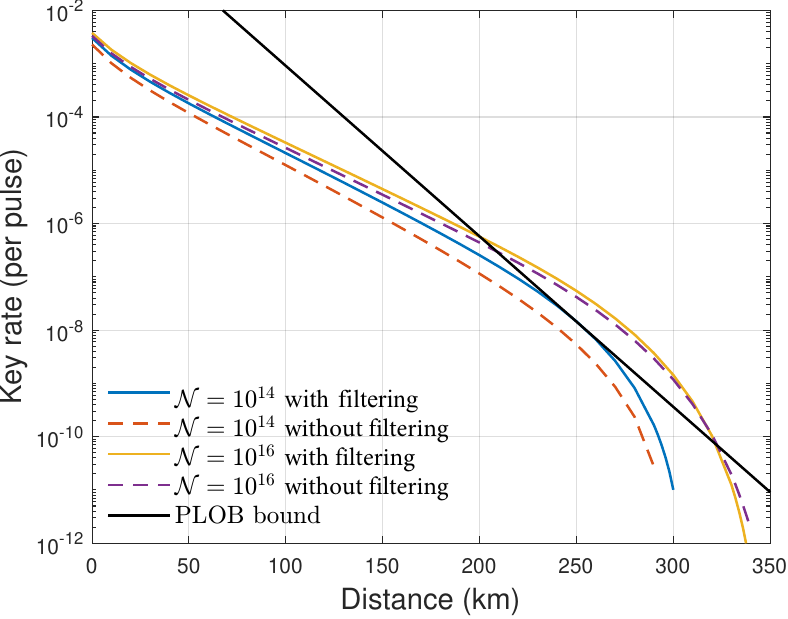}
\caption{Conference key rate of the AMDI-QCKA protocol in the finite-size regime for three users with three decoy states. Global phase locking is not used. We simulate the case where the total number of pulses each user sends as $10^{14}$, and $10^{16}$, with and without the click filtering method. The Chernoff bound~\cite{yin2020tight} is employed for finite-key analysis, and we set the failure probability $10^{-7}$. Our protocol can break the PLOB bound in finite size regime. 
}\label{fig_SKR_finite}
\end{figure}

In Fig.~\ref{fig_SKR_finite} we show the key rate of AMDI-QCKA in the finite size regime with three decoy states in the three-user case. Global phase-locking is not utilized.  The results show that even in finite size regime, our protocol still surpasses the PLOB bound at 200 km, and the maximum transmission distance exceeds 300 km. In the finite-key regime, 
a desired pairing event requires multiple users to simultaneously select the intended intensity. For instance, a $[2\nu,2\nu,2\nu]$ event requires all three users to choose the decoy intensity, with its probability of occurrence being approximately proportional to $p_{\nu}^6$.
To increase the data size,
we can use $[\mu,\mu,\nu]$, $[\nu,\mu,\mu]$, $[\mu,\nu,\mu]$, $[\mu,\nu,\nu]$, $[\nu,\mu,\nu]$, $[\nu,\nu,\mu]$ and $[\nu,\nu,\nu]$ events as $\boldsymbol{Z}$-basis events. Moreover, by employing the click filtering method~\cite{zhou2023experimental} to pre-exclude unwanted pairing events, such as the $[2\nu,2\nu,\nu+\mu]$ event, the key rate can be further enhanced. When using click filtering, after Eve announces the response events, the users disclose all time bins where they sent the decoy intensity 
$\nu$. When a time bin records $P_i$ clicks, and one of two users neighboring $P_i$ send intensity $\nu$ and another user sent 
$\mu$, such time bins are discarded. Note that when using click filtering, only $[\mu,\mu,\mu]$ 
events are used in $\boldsymbol{Z}$-basis.
In simulation we considered both cases, with and without click filtering method.

{
To experimentally realize the AMDI-QCKA protocol, high interference visibility is required. The maximum phase misalignment can be estimated as
$\delta = T_c \left(2\pi \Delta f + \omega_{\text{fiber}}\right) $,
where $\Delta f$ is the frequency mismatch between independent lasers and $\omega_{\text{fiber}}$ is the fiber-induced phase drift rate. 
Using the engineering criterion $\delta \lesssim 1$ rad to ensure acceptable interference visibility, the coherence time can be approximated as $T_c \approx \frac{1}{\omega_{\text{fiber}}}$ when $\Delta f$ is small (e.g., 10 Hz). With realistic parameters $\Delta f = 1~\text{kHz}$, $T_c = 50~\mu\text{s}$, $\omega_{\rm fiber}=3000$ rad/s and $M = 16$, we obtain $\delta \approx 0.46$, resulting in a $\sim 2\%$ increase in the X-basis QBER compared to the phase-locked case. These values are well within reach of current experimental techniques. 
Moreover, the key techniques required for our protocol have already been developed in previous MDI-QCKA implementations~\cite{yang2024experimental,du2024experimental} and asynchronous MDI-QKD systems~\cite{Zhu2023Experimental,zhou2023experimental,ZhangExperimental2025,shao2025mdiqc}. These advances indicate that an experimental demonstration of our protocol is within reach using existing technologies.
}

\section{Testing Mermin’s inequality} \label{sec_mermin}
The basis of the AMDI-QCKA protocol is that $N$ users share post-selected GHZ states, enabling a time-reversed GHZ experiment. In a traditional GHZ experiment, a GHZ state is created and distributed to distant observers. Each observer measures their own states in a randomly chosen basis. According to local realism, the measurement results should be predetermined and satisfy Mermin’s inequality~\cite{mermin1990simple}. The Mermin value can also be used to evaluate the quality of GHZ entanglement. When implementing the time-reversed GHZ experiment in the AMDI-QCKA network, the measurement step in a conventional GHZ test becomes state preparation, and the GHZ-entangled state is measured at the end of each experimental run instead of being prepared at the beginning. Specifically, the users randomly and independently send phase-randomized weak coherent pulses. They pair click events, and assign the pairing events to two complementary bases $\boldsymbol{X}$-basis and $\boldsymbol{Y}$-basis according to the phase reference frame. 
Using the detailed version of the key mapping rule in Sec.~\ref{sec_protocol}, the $\boldsymbol{X}$-basis is the pairing event with 
$\kappa_i=0$ or $1$, $\forall \vartheta_i$, and the $\boldsymbol{Y}$-basis is with $ \kappa_i=\frac{1}{2}$ or $\frac{3}{2}$, $\forall \vartheta_i$.
By applying the decoy state method, the post-selected GHZ states contributed by the single-photon components can be estimated. For the tripartite GHZ state $\ket{\Phi_0^+}$, 
{
Mermin’s inequality is}~\cite{mermin1990simple,fu2015long}
\begin{equation}
\begin{aligned} 
M_{111}^{\Phi_0^+} =& \langle XXX \rangle_{111}^{\Phi_0^+} - \langle XYY \rangle_{111}^{\Phi_0^+} \\
&- \langle YXY \rangle_{111}^{\Phi_0^+}- \langle YYX \rangle_{111}^{\Phi_0^+} \le 2,
\end{aligned}
\end{equation}
where $\langle XXX \rangle_{111}^{\Phi_0^+}$ is the mean value of $\ket{\Phi_0^+}$ state in the $\boldsymbol{X}$-basis, $M_{111}^{\Phi_0^+}$ is the Mermin value, with maximum value 4 for quantum system and 2 according from local realism. In Fig.~\ref{fig_merminvalue}, we simulate the $M_{111}^{\Phi_0^+}$ using formulas in Appendix~G and show that the post-selected GHZ states generated by our protocol can exceed 2. This demonstrates that our protocol enables a group of users in a network to share high-fidelity post-selected GHZ entanglement over long distances.


\begin{figure}[t]
\centering
\includegraphics[width=\linewidth]{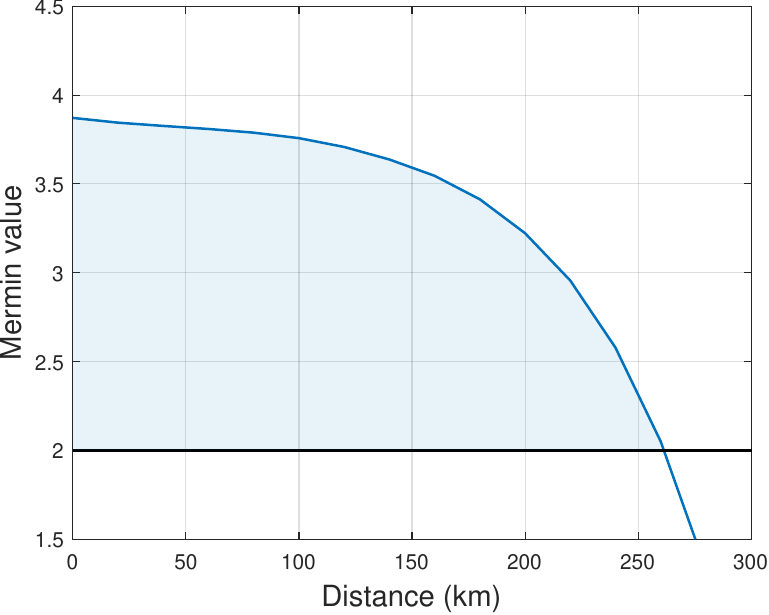}
\caption{The Mermin value $M_{111}^{\Phi_0^+}$ versus fiber channel transmission distance. We consider the finite-size scenario with $N = 10^{16}$. The decoy state method is used to estimate the single-photon components, and the result is compared with the classical bound of 2, the maximum value allowed by local realism.  
}\label{fig_merminvalue} 
\end{figure}

\section{Conclusion} \label{sec_conclude}
In this work, we propose an efficient measurement-device-independent quantum
conference key agreement protocol that surpasses the fundamental key rate limit in repeaterless quantum networks. 
By introducing asynchronous pairing of multiple detection events in a ring-interference network as coincidence events, our protocol enables the distribution of post-selected GHZ entanglement over long distances. The conference key rate is increased to scale as $O(\eta)$ at intercity distances, significantly improving efficiency. Our protocol achieves measurement-device independence by utilizing post-selected GHZ entanglement. 
Our approach utilizes weak coherent pulses and linear optics, while removing the requirement for global phase-locking, thereby greatly reducing the experiment complexity. The presented key rate formula with composable security incorporates the decoy state method, enhancing security. 
The use of a fixed number of decoy-state intensities boosts network scalability. Moreover, the violation of the Mermin inequality demonstrates the ability of our method to distribute post-selected GHZ states. In the future, by introducing $\boldsymbol{Y}$-basis measurement to more tightly estimate the phase error rate, the six-state encoding can be employed in AMDI-QCKA to further increase performance ~\cite{wang2023tight}. Since the techniques required for our protocol have already been developed in MDI-QCKA~\cite{fu2015long,yang2024experimental,du2024experimental} and asynchronous MDI (also called mode-pairing) quantum key distribution~\cite{xie2022breaking,zeng2022mode,Zhu2023Experimental,zhou2023experimental,zhu2024field,ge2024post,ZhangExperimental2025,shao2025mdiqc}, we anticipate that an experimental demonstration of our protocol can be realized in the near future using current experimental technologies. 

The core idea presented in this paper not only introduces new approaches for QCKA, but also has the potential to advance various quantum information tasks. By employing the $\boldsymbol{X}$-basis to extract secure keys and the $\boldsymbol{Z}$-basis to estimate the phase error rate, our scheme can be adapted for quantum secret sharing.  Moreover, the AMDI-QCKA network incorporates an interference loop spanning a two-dimensional plane, distinguishing it from a conventional one-dimensional single-chain quantum key distribution system.
This feature suggests potential applications in quantum sensing~\cite{lindsey2019illuminating, zhao2021field}. Additionally, the interference network of AMDI-QCKA provides insights into 
investigating quantum network coherence and network nonlocality~\cite{renou2019genuine,bibak2024quantum}. Finally, the ability of our protocol to distribute post-selected GHZ states provides numerous possibilities for developing novel quantum protocols, such as multi-party computing and quantum repeaters.

\section*{ACKNOWLEDGMENTS} 
This work is supported by the National Natural Science
Foundation of China (Grant No. 12274223), the Fundamental
Research Funds for the Central Universities and the
Research Funds of Renmin University of China (Grant No.
24XNKJ14), and the Program for Innovative Talents and
Entrepreneurs in Jiangsu (Grant No. JSSCRC2021484).


%

\end{document}


\appendix 

\title{Supplementary Material}

\author{Yu-Shuo Lu}\thanks{These authors contributed equally.}
\affiliation{National Laboratory of Solid State Microstructures and School of Physics, Collaborative Innovation Center of Advanced Microstrucstures, Nanjing University, Nanjing 210093, China}
\affiliation{School of Physics and Beijing Key Laboratory of Opto-electronic Functional Materials and Micro-nano Devices, Key Laboratory of Quantum State Construction and Manipulation (Ministry of Education), Renmin University of China, Beijing 100872, China}
\author{Hua-Lei Yin}\email{hlyin@ruc.edu.cn}\thanks{These authors contributed equally.}
\affiliation{School of Physics and Beijing Key Laboratory of Opto-electronic Functional Materials and Micro-nano Devices, Key Laboratory of Quantum State Construction and Manipulation (Ministry of Education), Renmin University of China, Beijing 100872, China}
\affiliation{National Laboratory of Solid State Microstructures and School of Physics, Collaborative Innovation Center of Advanced Microstrucstures, Nanjing University, Nanjing 210093, China}
\affiliation{Beijing Academy of Quantum Information Sciences, Beijing 100193, China}
\affiliation{Yunnan Key Laboratory for Quantum Information, Yunnan University, Kunming 650091, China}
\author{Yuan-Mei Xie}\thanks{These authors contributed equally.}
\affiliation{National Laboratory of Solid State Microstructures and School of Physics, Collaborative Innovation Center of Advanced Microstrucstures, Nanjing University, Nanjing 210093, China}
\affiliation{School of Physics and Beijing Key Laboratory of Opto-electronic Functional Materials and Micro-nano Devices, Key Laboratory of Quantum State Construction and Manipulation (Ministry of Education), Renmin University of China, Beijing 100872, China}
\author{Yao Fu}
\affiliation{Beijing National Laboratory for Condensed Matter Physics and Institute of Physics, Chinese Academy of Sciences, Beijing 100190, China}
\author{Zeng-Bing Chen}\email{zbchen@nju.edu.cn}
\affiliation{National Laboratory of Solid State Microstructures and School of Physics, Collaborative Innovation Center of Advanced Microstrucstures, Nanjing University, Nanjing 210093, China}

\maketitle

\begingroup
\color{black}
\tableofcontents
\endgroup

\section{Pairing algorithm} \label{sec_app_pair}
In this section, we introduce an example of the pairing algorithm for AMDI-QCKA, as shown in Algorithm~\ref{label_algorithm_pairing}. The pairing strategy is that $N$ nearest click events occurring at $N$ different ports are paired. There may exist more efficient pairing strategies. For security, the pairing strategy needs to meet two conditions~\cite{zhou2023experimental}. Firstly, from Eve's point of view, the pairing process should be independent of the basis. Secondly, the total intensity of the pair event is randomly decided, and is independent of Eve's operation.

\begin{algorithm} 
\DontPrintSemicolon
\SetAlgoLined
\caption{Pairing algorithm for the AMDI-QCKA}  
\KwIn{Detection result sequence $D = \{d_1,d_2,...\}$, where $d_i$ is the $i$-th detection event, with properties $T$ (the time of the click event), $P\in\{$$P_1,P_2,$ $...,P_N\}$ (the detection port) and $f=\{0,1\}$ (whether the click event has been paired).}
\KwOut{The set of pairing event  $\mathcal{P} = \{p_1, p_2,...\}$,  where  $p_i$ includes $N$ click events at $N$ different ports that are paired.}
\BlankLine
Initialize the pairing result sequence $\mathcal{P}$, set $Start\_index = 1$, 
$Repeat\_entry = 0$, $j = 1$;\;
\While{$\text{Start\_index} \leq \text{length}(D)$}{
    \For{$t = \text{Start\_index}:\text{length}(D)$}{
        \If{$D[t].T > D[\text{Start\_index}].T + T_c$\tcp*{Verify whether time interval is larger that $T_c$}}
        {
            Set the index of the last unused click event as $Start\_index$\;
            Break\;
        }        
        \If{$D[t].f = 0$}{
            $i = \text{Get\_DetectionPort}(D[t].P)$\;
            \uIf{
            $\mathcal{P}[j]$
              \text{\rm{lacks click event in} $P_i$}}{
               Assign the click event  $D[t]$ to pairing event $\mathcal{P}[j]$ \;
                $D[t].f = 1$\;
            }
            \ElseIf{Repeat\_flag $= 0$}{
                $Repeat\_entry = t$ \tcp*{Record the unused click event}
                $Repeat\_flag = 1$\;
            }
        }
        \If{$\mathcal{P}[j]$ \rm{include detection events of all ports}}
        {
            The $j$-th pairing event has been obtained\;
            $j = j + 1$\;
            \eIf{Repeat\_flag$ = 1$}{
                ${Start\_index} = {Repeat\_entry}$\;
                $Repeat\_flag = 0$
                }
            {
                Set $Start\_index$ to the last click event of $P[j]$\;
                Break\;
            }
        }
    }
} \label{label_algorithm_pairing}
\end{algorithm}  

\section{Security proof} \label{sec_app_security}
In this section, we prove the security of the AMDI-QCKA protocol. We provide the security proof by using the entanglement swapping argument. 
The basis of QCKA is that all legitimate users share almost perfect multiparty entanglement states. Assume $N$ users each prepare entangled state which contains a local qubit and an optical mode. They send the optical mode to untrusted relay Eve to perform the GHZ state measurement and post-select the successful GHZ state measurement events~\cite{fu2015long}, leaving the local qubits to be entangled.   
In the following we present the rigorous security proof and its equivalent protocol reduction. We start from virtual protocols, which can be reduced to the practical protocol described in the main text. 

\begin{figure}[t]
\includegraphics[width=0.49\textwidth]{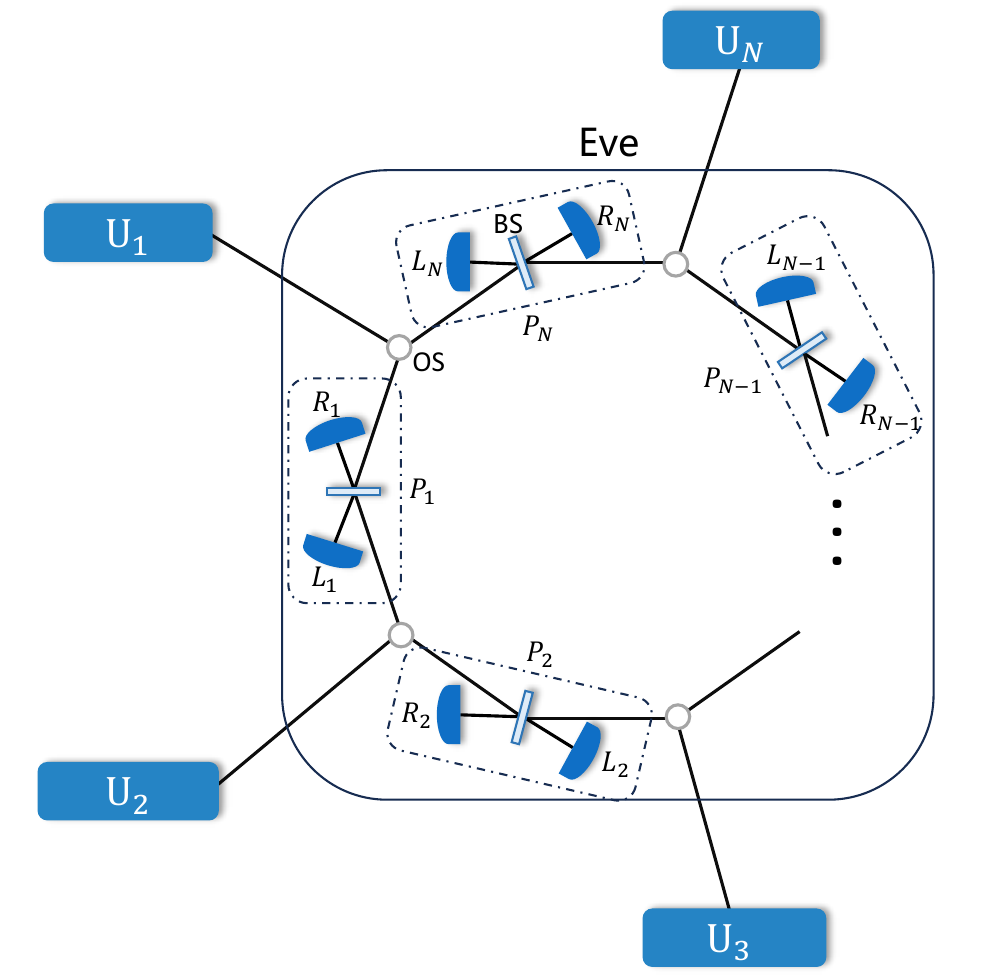}
\caption{The setup of Eve for the virtual protocol 1. The untrusted relay Eve is expected to consist of $N$ detection ports $P_1$, $P_2$, ..., $P_N$, and each port $P_i$ includes a 50/50 beam splitter, right detector $R_i$ and left detector $L_i$. Eve is expected to utilize optical switches (OS) to control the detection port where interference occurs. For the practical protocol in the main text, with the post-selection, we can replace the 
OS with $1\times2$ beam splitters.
}\label{fig_supp_vp1}
\end{figure}

\begin{figure}[t]
\includegraphics[width=0.5\textwidth]{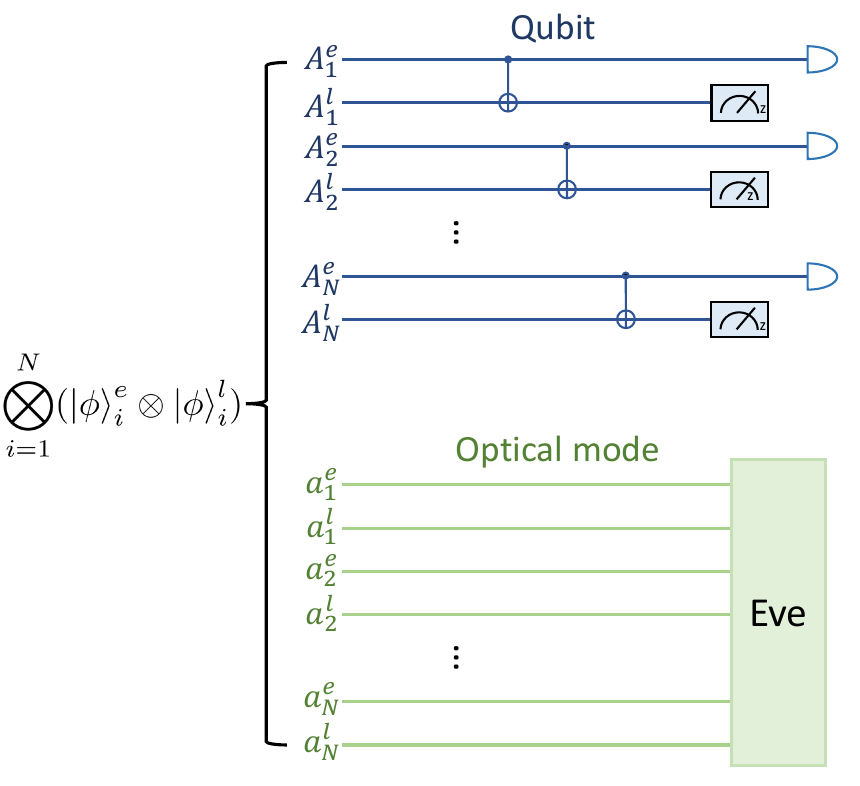}
\caption{ Schematic diagram of the entanglement-based virtual protocol 1 for the $N$-user asynchronous QCKA protocol. Each user prepares two entangled states $\ket{\phi}_{i}^{e(l)}=\frac{1}{\sqrt{2}}(\ket{+z}_{A_{i}^{e(l)}}\ket{1}_{a_{i}^{e(l)}} + 
e^{\mathbf{i}\theta_i}
\ket{-z}_{A_{i}^{e(l)}}\ket{0}_{a_{i}^{e(l)}})$. The users apply operator CNOT operation on $A_i^l$ controlled by $A_i^e$, and measure $A_i^l$ in the $\boldsymbol{Z}$-basis. Once the detection results of optical modes at Eve’s station constitute a coincidence event, the $N$-qubit GHZ state in the local qubit $\{A_1^e,A_2^e,...,A_N^e\}$ 
are obtained.
}\label{fig_supp_securitproof}
\end{figure}

\textbf{Virtual protocol 1.} (i)  As shown in Fig.~\ref{fig_supp_securitproof},
Each user $U_i$ ($i \in \{1,2, ...,N\}$) prepares two entangled states $\ket{\phi}_{i}^{e(l)}=\frac{1}{\sqrt{2}}(\ket{+z}_{A_{i}^{e(l)}}\ket{1}_{a_{i}^{e(l)}} + 
e^{\mathbf{i}\theta_i^{e(l)}}
\ket{-z}_{A_{i}^{e(l)}}\ket{0}_{a_{i}^{e(l)}})$,
where $\ket{\pm z}$ denote the two eigenstates of the Pauli $\boldsymbol{Z}$ operator $\sigma_{z}$, $\ket{0}$ and $\ket{1}$ are the zero and single-photon state, respectively, $\theta_i^{e(l)}$ is the relative phase, $A_i^{e(l)}$($a_i^{e(l)}$) denotes the local qubit (optical mode). 
{We also have two eigenstates in the $\boldsymbol{X}$-basis $\ket{\pm x}=(\ket{+z}\pm \ket{-z})/\sqrt{2}$}. The users pre-pair $N$ time bins. Each user applies a CNOT operation on $A_i^l$ controlled by $A_i^e$. They measure $A_i^l$ in the $\boldsymbol{Z}$-basis, and keep   $A_i^e$ if their measurement results are all $\ket{+z}$.  At time bin $i$,  $U_i$ and $U_{i+1}$ send optical modes $a_i^l$ and $a_{i+1}^e$ respectively to the untrusted relay, Eve (we define $\ket{\phi}_{N+1}^{e}  := \ket{\phi}_{1}^{e} $ to ensure a cyclic structure in the notation).  They repeat for multiple rounds to accumulate sufficient data. 
(ii) Eve is expected to perform single-photon interference measurement on incoming optical modes. At time bin $i$, $a_i$ interference with $a_{i+1}$ at detection port $P_i$, and  $a_N$ interference with $a_{1}$ at detection port $P_N$, as shown in Fig.~\ref{fig_supp_vp1}.
A successful click event is obtained when one and only one detector clicks. Eve announces the click time bin and the corresponding detector. 
(iii) For the detection result that Eve announces, the users post-select $N$-fold coincidence with $\theta_{g}^d=0$ or 1 (
$\theta_{g}^d := \sum\limits_{i=1}^{N} \theta_i^d$) as a successful event.
They identify the corresponding GHZ state according to the clicked detector.  Afterwards, the kept qubits $A_1^e$, $A_2^e$, .., $A_N^e$ establish entanglement.  (iv) Each user randomly chooses the $\boldsymbol{Z}$ and $\boldsymbol{X}$-bases to measure the qubits $A_1^e$, $A_2^e$, .., $A_N^e$ and obtains the corresponding bit values. 
Based on the detector click information, the users decide whether to flip the bit value.
(v) All the users publish the basis information through an authenticated classical channel. They use the data in the $\boldsymbol{X}$-basis to estimate the phase error rate, and the data in the $\boldsymbol{Z}$-basis to generate the conference key. By implementing a multipartite error correction and privacy amplification algorithm, they extract the final secret conference key. 

The quantum state evolution is given as follows. When all users prepare $\ket{\phi}_{i}^e$ and  $ \ket{\phi}_{i}^l$, the joint state is
\begin{widetext}
\begin{equation}
\begin{aligned}
\bigotimes\limits_{i=1}^N (\ket{\phi}_{i}^e \otimes \ket{\phi}_{i}^l)=&\bigotimes\limits_{i=1}^N  \left[\frac{1}{\sqrt{2}}\left(\ket{+z}_{A_{i}^e}\ket{1}_{a_{i}^e}+ e^{\mathbf{i}\theta_i^e}
\ket{-z}_{A_{i}^e}\ket{0}_{a_{i}^e}\right) \otimes \frac{1}{\sqrt{2}}\left(\ket{+z}_{A_{i}^l}\ket{1}_{a_{i}^l}+e^{\mathbf{i}\theta_i^l}\ket{-z}_{A_{i}^l}\ket{0}_{a_{i}^l}\right)\right]  \\
=&\bigotimes\limits_{i=1}^N
\frac{1}{2}
\left(
\ket{+z+z}_{A_{i}^eA_{i}^l}\ket{11}_{a_{i}^ea_{i}^l}+
e^{\mathbf{i}\theta_i^l}\ket{+z-z}_{A_{i}^eA_{i}^l}\ket{10}_{a_{i}^ea_{i}^l} +
e^{\mathbf{i}\theta_i^e}\ket{-z+z}_{A_{i}^eA_{i}^l}\ket{01}_{a_{i}^ea_{i}^l} \right.\\
&\left.+
e^{\mathbf{i}(\theta_i^l+\theta_i^e)} \ket{-z-z}_{A_{i}^eA_{i}^l}\ket{00}_{a_{i}^ea_{i}^l}
\right)  \\
\xrightarrow{\rm CNOT} &\bigotimes\limits_{i=1}^N
\frac{1}{2}
\left(
\ket{+z-z}_{A_{i}^eA_{i}^l}\ket{11}_{a_{i}^ea_{i}^l}+
e^{\mathbf{i}\theta_i^l}\ket{+z+z}_{A_{i}^eA_{i}^l}\ket{10}_{a_{i}^ea_{i}^l} +
e^{\mathbf{i}\theta_i^e}\ket{-z+z}_{A_{i}^eA_{i}^l}\ket{01}_{a_{i}^ea_{i}^l} \right.\\
& +\left. e^{\mathbf{i}(\theta_i^l+\theta_i^e)} \ket{-z-z}_{A_{i}^eA_{i}^l}\ket{00}_{a_{i}^ea_{i}^l}
\right).
\end{aligned}
\end{equation} 
\end{widetext}
After the users perform CNOT operation, measured $A_i^l$ and keep terms with $\ket{+z}$,  the joint state is 
\begin{equation}
\begin{aligned} 
&\bigotimes\limits_{i=1}^N
\frac{1}{\sqrt{2}}
\left[
\ket{+z}_{A_i^e}\ket{10}_{a_i^ea_i^l} +
e^{\mathbf{i}\theta_i^d}\ket{-z}_{A_i^e}\ket{01}_{a_i^ea_i^l}
\right]\\
&=\frac{1}{\sqrt{2}}\left[\ket{+x}_{A_i^e}
\frac{\ket{10}_{a_i^ea_i^l}+e^{\mathbf{i}\theta_i}\ket{01}_{a_i^ea_i^l}}{\sqrt{2}} \right.+ \left. \ket{-x}_{A_i^e}
\frac{\ket{10}_{a_i^ea_i^l}-e^{\mathbf{i}\theta_i^d}\ket{01}_{a_i^ea_i^l}}{\sqrt{2}}
\right],
\end{aligned}
\end{equation} 
where $\theta_i^d=\theta_i^e-\theta_i^l$ is the relative phase difference. They send their states to Eve, who utilizes an optical switch (OS) to control the path of the photon for interference measurement at each port, the joint state evolves as

\begin{equation}
\begin{aligned}
& \frac{1}{{2^{\frac{N}{2}}}}
\sum\limits_{{
\substack{c_1, c_2, ..., c_N \in \{0,1\} }}}^N 
\bigotimes_{i=1}^N  e^{\mathbf{i}(\theta_i^d)^{c_i}}
\left(\sigma_x^{c_i}\ket{+z}_{A_i}\right)
\ket{\overline{c}_ic_i}_{a_i^ea_i^l}  \\
\xrightarrow{\rm OS}& 
\frac{1}{{2^{\frac{N}{2}}}}
\sum\limits_{{
\substack{c_1, c_2, ... \\ c_N \in \{0,1\} }}}^N 
e^{\mathbf{i}\sum\limits_{i=1}^{N} (\theta_i^d)^{c_i}}\bigotimes_{i=1}^N
\left(\sigma_x^{c_i}\ket{+z}_{A_i}\right)
\ket{c_i\overline{c}_{i+1}}_{a_i^la_{i+1}^e}. 
\end{aligned}
\end{equation} 
Because only the $N$-fold coincident click events are kept, the terms with  $c_i\oplus \overline{c}_{i+1}=0$ are filtered out. The terms with $N$-fold coincidence is
\begin{equation}
\begin{aligned} 
&
\textcolor{red}{
\frac{1}{\sqrt{2}}}\left(
\ket{+z}^{\otimes N}
\bigotimes_{i=1}^N
\ket{01}_{a_i^la_{i+1}^e}  \!\! 
\!+\! e^{\mathbf{i}\theta_{g}^d} 
\ket{-z}^{\otimes N} 
\bigotimes_{i=1}^N
\ket{10}_{a_i^la_{i+1}^e} \!\!
\right).
\end{aligned}
\end{equation} 
When $\theta_{g}^d=0$ and $\pi$, the final joint states  are
\begin{equation}
\begin{aligned} \label{eq_virpro1_phi0+}
\frac{1}{{\sqrt{2}^N}}
\left(
\ket{+z}^{\otimes N} 
+ 
\ket{-z}^{\otimes N} 
\right)
\left(
\sum\limits_{\mathcal{S}^+}
\bigotimes_{i=1}^N 
\ket{r_i} 
\right),
\end{aligned}
\end{equation} 
and 
\begin{equation}
\begin{aligned} \label{eq_virpro1_phi0-}
\frac{1}{{\sqrt{2}^N}}\left(
\ket{+z}^{\otimes N} 
-
\ket{-z}^{\otimes N} 
\right)
\left(
\sum\limits_{\mathcal{S}^-}
\bigotimes_{i=1}^N 
\ket{r_i} 
\right),
\end{aligned}
\end{equation} 
where the summation constraint are $\mathcal{S^+}: r_1 \oplus r_2 \oplus ... \oplus r_N =
(N-1~\rm {mod}~2)$ and $\mathcal{S^-}: r_1 \oplus r_2 \oplus ... \oplus r_N =
(N~\rm {mod}~2)$, $r_i \in \{0,1\}$ corresponding to detector $\{L_i,R_i\}$ clicks, respectively. With the detector click result given in Eq.~\eqref{eq_virpro1_phi0+} and \eqref{eq_virpro1_phi0-}, the local qubits
$\{A_1^e, A_2^e, ..., A_N^e\}$  establish GHZ entanglement $\ket{\Phi_0^+}$ and $\ket{\Phi_0^-}$, respectively.

In virtual protocol 1, Eve cannot eavesdrop the secret keys without being found. 
The measurement of qubit $A_i^e$ is performed locally by the users, and $\boldsymbol{X}$ and $\boldsymbol{Z}$-bases are randomly chosen and unknown to Eve before Eve broadcasts the click information, and Eve's operation will give rise to phase error rate. Once the basis choice information is unknown to Eve, the security can be guaranteed.  
Thus the pre-pairing of virtual protocol 1 is secure.

In addition, the measurement of $A_i^e$ can be preceded to step (i) because the measurement operation commutes other steps. Thus the entanglement-based protocol can be equivalently transferred to prepare-and-measure protocol. When the measurement outcome of $A_i^e$ is $\ket{+z}$ ($\ket{-z}$), $U_i$ prepares the state $\ket{10}_{a_i^ea_i^l}$ ($\ket{01}_{a_i^ea_i^l}$) . When the outcome is $\ket{+x}$ ($\ket{-x}$), $U_i$ prepares $\frac{1}{\sqrt{2}}(\ket{10}_{a_i^ea_i^l}+e^{\mathbf{i}\theta_i^d}\ket{01}_{a_i^ea_i^l})$ ($\frac{1}{\sqrt{2}}(\ket{10}_{a_i^ea_i^l}-e^{\mathbf{i}\theta_i^d}\ket{01}_{a_i^ea_i^l})$). Then we have the virtual protocol 2, where the users directly prepare optical modes instead of entangled states.

\textbf{Virtual protocol 2.} 
The $N$ users pre-pair $N$ time bins. Each user $U_i$ randomly chooses  $\boldsymbol{X}$ or $\boldsymbol{Z}$-basis. If the users chooses $\boldsymbol{Z}$-basis, they randomly prepare optical modes $\ket{10}_{a_i^ea_i^l}$ and $\ket{01}_{a_i^ea_i^l}$. If they choose $\boldsymbol{X}$-basis, they prepares state $
\textcolor{red}{
\frac{1}{\sqrt{2}}}(\ket{10}_{a_i^ea_i^l}\pm e^{\mathbf{i}\theta_i^d}\ket{01}_{a_i^ea_i^l})$.  $U_i$ and $U_{i+1}$ send $a_i^l$ and $a_{i+1}^e$ respectively to the measurement node at time bin $i$. The following steps, including interference measurement, parameter estimation, error correction and privacy amplification are the same as virtual protocol 1. 

In real implementation, the phase-randomized coherent source can be utilized as replacement because of the difficulty and low efficiency in directly preparing single-photon states. With a random global phase of the coherent pulse, from the view of Eve, the joint state that each user send is Fock state. Using decoy state method~\cite{wang2005beating,lo2005decoy} and tagged model~\cite{gottesman2004security}, one can effectively estimate the amount of  single-photon events and obtain secure secret key.  Therefore we have virtual protocol 3, where the users prepare phase-randomized coherent state instead of single-photon state. 
 
\textbf{Virtual protocol 3.} 
The $N$ users pre-pair $N$ time bins. Each user $U_i$ randomly chooses $\boldsymbol{X}$ or $\boldsymbol{Z}$-basis with probabilities $p_z$ and $p_x$, respectively. They randomly prepare quantum states$\ket{e^{\mathbf{i}\theta_i^e}\sqrt{k_i}}_{a_i^e}\ket{0}_{a_i^l}$ and $\ket{0}_{a_i^e}\ket{e^{\mathbf{i}\theta_i^l}\sqrt{k_i}}_{a_i^l}$ in the $\boldsymbol{Z}$-basis, and  $ \ket{e^{\mathbf{i}\theta_i^e}\sqrt{k_i}}_{a_i^e}   \ket{e^{\mathbf{i}\theta_i^l} \sqrt{k_i}}_{a_i^l}$ in the $\boldsymbol{X}$-basis. The following steps, are the same as virtual protocol 2. 

There are two differences between the virtual protocol 3 and the practical protocol in the main text. First, the pairing information is pre-determined by the users 
{in virtual protocol 3}, but in the practical protocol it is determined by the detection information announced by Eve. In both cases each user sends a train of coherent states with random phase and random intensity from the view of Eve. Eve obtains same classical and quantum information in two protocols. The random basis and the bit information is determined by the intensity of the pairing event and the phase difference between two paired optical pulses. Even if Eve can control the pairing result, the bit information is still inaccessible, because the modulated phase information is stored in the user's local classical memory. In addition, as long as the pairing result is not related with the basis information, Eve can not obtain extra basis information. Finally, 
guaranteed that the total intensity of the pairing event is randomly decided, the single-photon components is uniformly distributed in the pairing events. Any attacks from Eve will be discovered with increased phase error rate in single-photon components. 
Therefore, the users can always utilize post-pair method to increase the valid data, with security level equal to the virtual protocol.

In addition, in virtual protocol 3, Eve is expected to actively control the detection port where interference occurs at each time bin. In practice, Eve's operation is not restricted. Eve can arbitrarily control the detection port and declare the measurement result. However, this inevitably gives rise to phase error rate. Therefore, we can adopt the post-selection strategy. All the users simultaneously send phase-randomized coherent state. When single click event occurs at detection port $P_i$, one can post-select this click event as a result of interference of $a_i^l$ and $a_{i+1}^e$, 
corresponding to the practical protocol.  

In conclusion, we have proven that the security of the AMDI-QCKA protocol in the main text is equivalent to that of the entanglement-based protocol.  
\section{Composable security} 
\label{sec_app_compsecu}
Here we prove the composable secutiy of the AMDI-QCKA protocol~\cite{li2023breaking,li2023breakingqss}. We first give the definitions of the security criteria for the QCKA. A QCKA protocol is $\varepsilon_{\rm cor}$-correct if~
\begin{equation}
\begin{aligned}
{\rm Pr}(\exists~i \in \{2,...,n\}, {\rm s.t.}~\textbf{S}_1 \neq \textbf{S}_i) \leq \varepsilon_{\rm cor},
\end{aligned}
\end{equation}
where $\textbf{S}_i$ ($i\in\{1,2,...,n\}$) is the final key of $U_i$. A QCKA protocol is called $\varepsilon_{\rm sec}$-secret if 
\begin{equation}
\begin{aligned}
p_{\rm pass}D\left(\rho_{\mathbf{S_1},E},\sum\limits_{\mathbf{S_1}}\frac{1}{|\mathbf{S_1}|}\ket{\mathbf{S_1}}\bra{\mathbf{S_1}}\otimes \sigma_E\right)\le \varepsilon_{\rm sec},
\end{aligned}
\end{equation}
where $p_{\rm pass}$ is the probability of passing the protocol, $E$ is the system of Eve,
$D(\cdot,\cdot)$ is the trace distance, $\sigma_E$ is the system of the eavesdropper.  

Then, we provide the composable security proof for the AMDI-QCKA protocol and the final key rate formula. 
 
\textbf{Theorem 1.} \emph{The AMDI-QCKA protocol is $\varepsilon_{\rm cor}$-correct.}

\emph{Proof:} Let $\{{\rm{EC}}_i\}_{i=2}^{n}$ be a set of error correction (EC) protocols, $P_{XK}$ is a probability distribution, where where $k_i$ is the guess of $x$ of the $i$-th user. Any set of EC protocols, which is $\varepsilon_{\rm cor}$-secure, satisfies $\mathrm{Pr}(\exists~i \in \{2,...,n\}, \mathrm{s.t.} x \neq  {k}_i) \leq\varepsilon_{\rm cor}$.
According to Theorem 2 in Ref.~\cite{grasselli2018finite}, given $P_{XK}$, 
there exists a one-way EC protocol with $\varepsilon_{\rm cor}$-fully secure on $P_{XK}$ with information leakage $ {\rm leak_{EC}}+\log_2\frac{2(N-1)}{\varepsilon_{\rm cor}}$.  
Therefore, the $\varepsilon_{\rm cor}$-correctnesss of the AMDI-QCKA protocol can be guaranteed during the error correction process.

\textbf{Theorem 2.} \emph{The QCKA protocol defined in the main text is $\varepsilon_{\rm sec}$-secret if the key length l satisfies Eq.(2) in the main text}. 

\emph{Proof.} The quantum leftover hashing lemma~\cite{PhysRevLett.81.3018} shows that if $U_1$ map the raw key $\mathcal{Z}$ to the final key $\mathbf{S}$ and extracts a string of length $l$ utilizing a random universal$_2$ hash function, for any positive $\epsilon$, we have
\begin{equation}
\begin{aligned}
\varepsilon_{\rm sec}
\ge \frac{1}{2}
\sqrt{2^{l-H_{\rm min}^\varepsilon(\mathcal{Z}|\textbf{E}')}} + 2\varepsilon,
\end{aligned}
\end{equation}
where $\textbf{E}'$ is the auxiliary system of Eve after error correction, and $H_{\rm min}^\varepsilon(\mathcal{Z}|\textbf{E}')$ can be bounded by 
\begin{equation}
\begin{aligned}
H_{\rm min}^\epsilon(\mathcal{Z}|\textbf{E}')
\ge H_{\rm min}^\epsilon(\mathcal{Z}|\textbf{E}) -{\rm leak_{EC}}-\log_2\frac{2(N-1)}{\varepsilon_{\rm cor}},
\end{aligned}
\end{equation}
where $\textbf{E}$ is the auxiliary system of Eve before error correction. 

To bound the term $H^\epsilon_{\rm min}(\mathcal{Z}|\textbf{E})$, using the chain-rule inequality for smooth entropies~\cite{Vitanov2013chain,zhou2023experimental}, we have 
\begin{equation}
\begin{aligned}
H^\varepsilon_{\rm min}(\mathcal{Z}|\textbf{E}) &\ge H_{\rm min}^{\varepsilon'+2\varepsilon_e+(\hat{\varepsilon}+2\hat{\varepsilon}'+\hat{\varepsilon}'')}
(\mathcal{Z}_0\mathcal{Z}_{\{1\}^N}\mathcal{Z}_{\rm rest}|\textbf{E}) \\
&\ge s_{0}^z + H_{\rm min}^{\varepsilon_e}
(\mathcal{Z}_{\{1\}^N}|\mathcal{Z}_{0}\mathcal{Z}_{\rm rest}\textbf{E}) 
\!-2\log_2(2/\varepsilon'\hat{\varepsilon}),
\end{aligned}
\end{equation}
where $\varepsilon =\varepsilon'+2\varepsilon_e+(\hat{\varepsilon}+2\hat{\varepsilon}'+ \hat{\varepsilon}'') $, $\mathcal{Z}_0$, $\mathcal{Z}_{\{1\}^N}$ and $\mathcal{Z}_{\rm rest}$ represents the bits when $U_1$ send vacuum state, the $N$ users each sends a single-photon state and the rest cases, respectively. We used the fact that $H_{\rm min}^{\hat{\varepsilon}'}(\mathcal{Z_{\rm rest}}|\mathcal{Z}_0\textbf{E})\ge 0$ and $H_{\rm min}^{\hat{\varepsilon}''}(\mathcal{Z}_0|\textbf{E})\ge H_{\rm min}(\mathcal{Z}_0) = s_{0}^z$.
Because the the single-photon component prepared in the $\boldsymbol{Z}$-basis and $\boldsymbol{X}$-basis are mutually unbiased, we can use phase error rate $ \phi_{\{1\}^{N}}^z$ to quantify the minimum smooth entropy using the entropic uncertainty relations~\cite{tomamichel2012tight,curty2014finite}:
\begin{equation}
\begin{aligned}
H_{\rm min}^{\varepsilon_e} (\mathcal{Z}_{\{1\}^N}|\mathcal{Z}_{0}\mathcal{Z}_{\rm rest}\textbf{E}) & \ge s_{\{1\}^N}^z  -
H_{\rm \max}^{\varepsilon_e}(\mathcal{X}_{\{1\}^N}^1|\mathcal{X}_{\{1\}^N}^2...\mathcal{X}_{\{1\}^N}^N) \\
&
\ge s_{\{1\}^N}^z[1-H_2( \phi_{\{1\}^{N}}^z)],
\end{aligned}
\end{equation}
where $\mathcal{X}_{\{1\}^N}^i$ represents the $\boldsymbol{X}$-basis string of  $U_i$  when the users hypothetically measure the single-photon state in the $\boldsymbol{X}$-basis instead of $\boldsymbol{Z}$-basis when each user sends a single-photon state.
Finally, by choosing $\hat{\varepsilon}' = \hat{\varepsilon}'' = 0$, we have the final key rate formula:
\begin{equation}
\begin{aligned}
l \ge & s_{0}^z + s_{\{1\}^N}^z[1-H_2( \phi_{\{1\}^{N}}^z)] - {\rm leak}_{\rm EC} -\log_2 \frac{2(N-1)}{\varepsilon_{\rm cor}} -2\log_2\frac{2}{\varepsilon'\hat{\varepsilon}}
-2\log_2\frac{1}{2{\varepsilon_{\rm PA}}},
\end{aligned}
\end{equation} 
where the overall security $\varepsilon_{\rm tot}$ is defined as: $\varepsilon_{\rm tot} = \varepsilon_{\rm sec}+ \varepsilon_{\rm cor}$,  $\varepsilon_{\rm sec} = 2(\varepsilon'+2\varepsilon_e+\hat{\varepsilon})+\varepsilon_0+\varepsilon_N + \varepsilon_\beta + \varepsilon_{\rm PA}$,  $\varepsilon_0$, $\varepsilon_N$ and $\varepsilon_\beta$ are the failure probabilities in estimating the terms of $s_{0}^z$, $s_{\{1\}^{N}}^z$, and $ \phi_{\{1\}^{N}}^z$. Specially, in finite case with three users and three decoy states, we have $\varepsilon_0 = 4\epsilon$, $\varepsilon_3 = 15\epsilon$ with $\epsilon$ is the failure probability of Chernoff bound. In simulation, we set all security parameters to the same value $\epsilon = 10^{-7}$.
 
\section{Decoy state estimation}
\label{sec_app_decoy}
\subsection{\texorpdfstring{$N$}{N} users with infinite decoys}
To ensure the security of the the AMDI-QCKA protocol, we employ the decoy state method. Secure secret keys can be distilled from the single-photon components.  
Initially, we address the general scenario where there are infinite number of decoy states with $N$ users. Subsequently, we present detailed formula for the specific 3 and 4-user cases. Finally, the three-decoy state formulas with three users are provided. 

With infinite number of decoy states, we assume the number of single-photon components can be accurately estimated. In the following, we focus on symmetric case, where the users randomly choose intensity from the same setting, the $ [\mu,\mu,...,\mu]$ events are assigned to $\boldsymbol{Z}$-basis, $[2\nu,2\nu,...,2\nu]$ events are assigned to $\boldsymbol{X}$-basis, and other events are used in parameter estimation. The final key length formula is
\begin{equation}\label{eq_keyrate_infdecoy}
\begin{aligned} 
l_{\rm asy} = &  s_{\{1\}^{N}}^z\left[1-H_2(\phi_{\{1\}^{N}}^z)\right]- {\rm leak}_{\rm EC},
\end{aligned}
\end{equation} 
where $s_{\{1\}^{N}}^{z(x)} $ is  the number of single-photon 
components in the $\boldsymbol{Z}$($\boldsymbol{X}$)-basis,  $\phi_{\{1\}^{N}}^z$ denotes the phase error rate,  
${\rm leak}_{\rm EC} =\max \limits_{2\le i \le N}[H_2(E^z_{1,i})]fn^z$, $n^{z}$ is the number of the $\boldsymbol{Z}$-basis event. $s_{\{1\}^{N}}^z$ can be given as
\begin{equation}
\begin{aligned} \label{eq_Y_nz}
s_{\{1\}^{N}}^z &= n_{\rm tot}
\left({p_\mu p_o  \mu e^{-\mu}  p_{\rm nc}} \right)^N
\sum\limits_{\substack{a_i^e+a_i^l=1\\
\forall i\in \{1,...,N\}}} \prod\limits_{i=1}^N 
\frac{y_{a_{i+1}^e a_{i}^l}^{P_i}}{q_{\rm tot}^{P_i}},
\end{aligned}
\end{equation} 
where $p_{\rm nc} = \left\{\left[1-(1-e^{-\overline{\mu}\eta})\right](1-p_d)^2\right\}^{N-1}$ represents the other $N-1$ ports have no click, $\overline{\mu}$ is the mean photon number from each user, $\eta$ is the transmittance from user to Eve, $n_{\rm tot}$ is the number of pairing event given in Eq.~\eqref{eq_ppair}, 
$a_i^{e(l)} \in \{0,1 \}$ is the number of incident photon of the early (late) time of $U_i$, $q_{\rm tot}^{P_i}$ is the probability that detection port $P_i$ have click. $y_{a_{i+1}^e a_{i}^l}^{P_i}$ is the probability of having a click in detection port $P_i$ when $a_{i+1}^e$ and $a_{i}^l$ photons from the two input paths of the beam splitter. To ensure a cyclic structure
in the notation, we define $a_{N+1}^{e(l)} :=a_{1}^{e(l)}$. For the $N$-user cases, there are $2^N$ cases of photon number distribution in the $\boldsymbol{Z}$-basis. In symmetric channels, the loss from each user to Eve are equal, thus we assume $y_{ab} = y_{a_{2}^e a_{1}^l}^{P_1}=y_{a_{3}^e a_{2}^l}^{P_i}=...=y_{a_{1}^e a_{N}^l}^{P_N}$ ($a,b\in\{0,1\}$). Therefore, for each possible $y_{ab}$ we have 
\begin{equation}
\begin{aligned}
y_{11} =& [2p_d(1-\eta')^2 + 2\eta'(1-\eta) + \frac{\eta'^2}{2}](1-p_d), \\ 
y_{10} =& y_{01} = [2p_d(1-\eta') + \eta'](1-p_d), \\
y_{00} =& 2p_d(1-p_d), 
\end{aligned}
\end{equation} 
where $p_d$ is the dark counting rate, $\eta' = \eta/2$, and the factor $1/2$ is introduce from the $1\times 2$ beam splitter. In the $\boldsymbol{X}$-basis, because the density matrices for the single-photon components in the two bases are the same,   the expected ratio of different intensity settings for single-photon components have the following relation:
\begin{equation}
\begin{aligned}
\frac{s_{\{1\}^{N}}^z}{s_{\{1\}^{N}}^x} = \frac{\mu^Ne^{-N\mu}p_{\vec{K_z}}}{(2\nu)^Ne^{-2N\nu}p_{\vec{K_x}}
},
\end{aligned}
\end{equation} 
where $\vec{K_z} = [\mu,\mu,...\mu]$, $ \vec{K_x}=[2\nu,2\nu,...,2\nu]$, 
\begin{equation}
\begin{aligned}
p_{[k_{1}^{tot},k_{2}^{tot},...,k_{N}^{tot}]}=
\sum\limits_{\substack{k_{i}^{e}+k_{i}^{l}= k_{i}^{\rm tot}\\
\forall i\in \{1,...,N\}}}\prod\limits_{i=1}^N  p_{k_{i}^{e}}p_{k_{i}^{l}},
\end{aligned}
\end{equation} 
except for 
$p_{\vec{K_x}}=\frac{2p_{\nu}^{2N}}{M}$. 
$\frac{2}{M}$ comes from the pairing condition $\theta_{g}^d/\pi \mod 2 =0$ or $1$.  
To obtain the phase error rate $ \phi_{\{1\}^{N}}^z$, we first calculate $e_{\{1\}^{N}}^x$, the bit error rate of the single-photon component of the $\boldsymbol{X}$-basis. In asymptotic limit we have $ \phi_{\{1\}^{N}}^z = e_{\{1\}^{N}}^x$, where $e_{\{1\}^{N}}^x $ can be calculated 
by using $ e_{\{1\}^{N}}^x =  t_{\{1\}^{N}}^x/s_{\{1\}^{N}}^x
$, and the number of error event is 
\begin{equation}
\begin{aligned}
t_{\{1\}^{N}}^x
=& 
n_{\rm tot}
\left(\frac{2}{M}\right)
\left({p_\nu^22\nu e^{-2\nu}}\right)^N \times
\frac{1}{\prod\limits_{i=1}^Nq_{\rm tot}^{P_i}}
\left[
(1-e_d) \mathcal{Y}^{\rm err}_{\{1\}^{N}}+
e_d\mathcal{Y} ^{\rm cor}_{\{1\}^{N}}
\right],
\end{aligned}
\end{equation} 
where $e_d$ is the misalignment error of $\boldsymbol{X}$-basis, $\mathcal{Y}^{ \rm err}_{\{1\}^{N}}$ ($\mathcal{Y} ^{\rm cor}_{\{1\}^{N}}$) is the probability of obtaining an error (correct) coincident event given single-photon components as input, which vary with the number of users. 
In the following, we introduce the detailed formulas for the specific three-user and four-user cases.

\subsection{\texorpdfstring{$N=3$}{N=3} with infinite decoy states}

For the case of $N$ = 3, according from Eq.~\eqref{eq_keyrate_infdecoy}, the key rate formula is $
l_{\rm asy} =s_{111}^z \left[1-H_2(\phi_{111}^z)\right]-{\rm leak}_{\rm EC}$.   
In the $\boldsymbol{Z}$-basis, by using Eq.~\eqref{eq_Y_nz}, $s_{111}^z$ can be given as:
\begin{equation}
\begin{aligned}
s_{111}^z =n_{\rm tot}\frac{(p_\mu p_o \mu e^{-\mu}p_{\rm nc})^3}
{q_{\rm tot}^{P_1}q_{\rm tot}^{P_2}q_{\rm tot}^{P_3}}
(2y_{10}y_{10}y_{10} + 6y_{11}y_{10}y_{00}).
\end{aligned}
\end{equation} 
The number of single-photon components in the $\boldsymbol{X}$-basis is calculated using the relation between $s_{111}^z$ and $s_{111}^x$:
$ 
\frac{s_{111}^z}{s_{111}^x} = \frac{\mu^3e^{-3\mu}p_{[\mu,\mu,\mu]}}{8\nu^3e^{-6\nu}p_{[2\nu,2\nu,2\nu]}}
$,
where $p_{[\mu,\mu,\mu]}=8(p_{\mu}p_o)^3$, $p_{[2\nu,2\nu,2\nu]}=\frac{2p_{\nu}^6}{M}$. We use the GHZ state $\ket{\Phi_0^+}$ as the reference state:
\begin{equation}
\begin{aligned}
\ket{\Phi_0^+}=& \frac{1}{\sqrt{2}}(\ket{+z+z+z}+\ket{-z-z-z}) \\
=& \frac{1}{2}(\ket{+x+x+x}+ \ket{+x-x-x}  + \ket{-x+x-x} + \ket{-x-x+x}).
\end{aligned}
\end{equation} 
When the click detectors of $\{P_1,P_2,P_3\}$ are $\{L_1,L_2,L_3\}$, $\{L_1,R_2,R_3\}$, $\{R_1,L_2,R_3\}$ or $\{R_1,R_2,L_3\}$, we obtain a  measurement event of  $\ket{\Phi_0^+}$. The other four cases corresponding to error event. Therefore, the number of error event of the single-photon state in the $\boldsymbol{X}$-basis can be given as   
\begin{equation}
\begin{aligned}
t_{111}^x &=
n_{\rm tot}\left(\frac{2}{M}\right)
\frac{(2\nu e^{-2\nu}p_\nu^{2})^3}{q_{\rm tot}^{P_1}q_{\rm tot}^{P_2}q_{\rm tot}^{P_3}}
\left[(1-e_d) 
\mathcal{Y}^{\rm err}_{111}  +  e_d \mathcal{Y}^{\rm cor}_{111}  \right],
\end{aligned}
\end{equation} 
where $\mathcal{Y}^{\rm err}_{111} = {y^{R,R,R}}+{y^{R,L,L}}+{y^{L,R,L}}+{y^{L,L,R}}$,  $\mathcal{Y}^{\rm cor}_{111} = {y^{L,L,L}}+{y^{L,R,R}}+{y^{R,L,R}+{y^{R,R,L}}}$,
$y^{R(L),R(L),R(L)}$ is the probability of having detector $R_1(L_1)$, $R_2(L_2)$, $R_3(L_3)$   click when input single-photon states in the $\boldsymbol{X}$-basis. Using the method derived in~\cite{ma2012alternative}, they can be calculated as follows. The input single-photon state can be written as
\begin{widetext}
\begin{equation}
\begin{aligned}
&\left[\frac{1}{\sqrt{2}}
(\ket{10}_{a_1^ea_1^l} + e^{\mathbf{i}\theta_1^d}\ket{01}_{a_1^ea_1^l})\right]
\otimes\left[\frac{1}{\sqrt{2}}(\ket{10}_{a_2^ea_2^l} + e^{\mathbf{i}\theta_2^d}\ket{01}_{a_2^ea_2^l})\right]
\otimes\left[\frac{1}{\sqrt{2}}(\ket{10}_{a_3^ea_3^l}+ e^{\mathbf{i}\theta_3^d}\ket{01}_{a_3^ea_3^l})\right]. 
\end{aligned}
\end{equation} 
After passing through channel with transmittance $\eta$, the state becomes a mixed state:   
\begin{equation}
\begin{aligned} \label{eq_3eta_psi}
&\frac{\eta^3}{8}\ket{\psi_{111}} \bra{\psi_{111}} + \frac{\eta^2(1-\eta)}{4}\left(\ket{\psi_{110}} \bra{\psi_{110}} + \ket{\psi_{101}} \bra{\psi_{101}} + \ket{\psi_{011}} \bra{\psi_{011}}\right) \\
& + \frac{\eta(1-\eta)^2}{2}\left(\ket{\psi_{100}} \bra{\psi_{100}} + \ket{\psi_{010}} \bra{\psi_{010}} + \ket{\psi_{001}} \bra{\psi_{001}}\right) +(1-\eta)^3\ket{\psi_{000}} \bra{\psi_{000}},
\end{aligned}
\end{equation}
where we use the state $\ket{a_1^e a_1^l a_2^e a_2^l a_3^e a_3^l}$ to represent photon number from different users' early (late) time bin of the single-photon state, and
\begin{equation}
\begin{aligned} 
\ket{\psi_{111}}_{a_1^e a_1^l a_2^e a_2^l a_3^e a_3^l}&= \ket{101010}+e^{\mathbf{i}\theta_3^d}\ket{101001}+e^{\mathbf{i}\theta_2^d}\ket{100110}+e^{\mathbf{i}(\theta_2^d+\theta_3^d)}\ket{100101}+e^{\mathbf{i}\theta_1^d}\ket{011010}\\
&+e^{\mathbf{i}(\theta_1^d+\theta_3^d)}\ket{011001}+e^{\mathbf{i}(\theta_1^d+\theta_2^d)}\ket{010110}+e^{\mathbf{i}(\theta_1^d+\theta_2^d+\theta_3^d)}\ket{010101} ,
\\
\ket{\psi_{110}}_{a_1^e a_1^l a_2^e a_2^l a_3^e a_3^l} &= \ket{101000}+e^{\mathbf{i}\theta_2^d}\ket{100100}+e^{\mathbf{i}\theta_1^d}\ket{011000}+e^{\mathbf{i}(\theta_1^d+\theta_2^d)}\ket{010100},
\\
\ket{\psi_{101}}_{a_1^e a_1^l a_2^e a_2^l a_3^e a_3^l} &= \ket{100010}+e^{\mathbf{i}\theta_3^d}\ket{100001}+e^{\mathbf{i}\theta_1^d}\ket{010010}+e^{\mathbf{i}(\theta_1^d+\theta_3^d)}\ket{010001},
\\
\ket{\psi_{011}}_{a_1^e a_1^l a_2^e a_2^l a_3^e a_3^l} &= \ket{001010}+e^{\mathbf{i}\theta_3^d}\ket{001001}+e^{\mathbf{i}\theta_2^d}\ket{000110}+e^{\mathbf{i}(\theta_2^d+\theta_3^d)}\ket{000101},
\\
\ket{\psi_{100}}_{a_1^e a_1^l a_2^e a_2^l a_3^e a_3^l} &= \ket{100000} + e^{\mathbf{i}\theta_1^d}\ket{010000}, \\
\ket{\psi_{010}}_{a_1^e a_1^l a_2^e a_2^l a_3^e a_3^l} &= \ket{001000} + e^{\mathbf{i}\theta_2^d}\ket{000100},\\
\ket{\psi_{001}}_{a_1^e a_1^l a_2^e a_2^l a_3^e a_3^l} &= \ket{000010} + e^{\mathbf{i}\theta_3^d}\ket{000001},\\
\ket{\psi_{000}}_{a_1^e a_1^l a_2^e a_2^l a_3^e a_3^l} &= \ket{000000}.
\end{aligned}
\end{equation} 
At the detection port $P_i$, the photon $a_i^l$ interferes with $a_{i+1}^e$. Therefore, we can apply the following state transformation of the beam splitter (BS) where: 
$\ket{0}_{a_i^l}\ket{0}_{a_{i+1}^e} \rightarrow \ket{0}_{L_{i}}\ket{0}_{R_{i}}$, $  \ket{1}_{a_i^l}\ket{1}_{a_{i+1}^e} \rightarrow (\ket{2}_{L_i}\ket{0}_{R_i} -\ket{0}_{L_i}\ket{2}_{R_i})/\sqrt{2}$,  $
\ket{1}_{a_i^l}\ket{0}_{a_{i+1}^e} \rightarrow (\ket{1}_{L_i}\ket{0}_{R_i}-\ket{0}_{L_i}\ket{1}_{R_i})/\sqrt{2}$,$ 
\ket{0}_{a_i^l}\ket{1}_{a_{i+1}^e} \rightarrow (\ket{1}_{L_i}\ket{0}_{R_i}+\ket{0}_{L_i}\ket{1}_{R_i})/\sqrt{2}$,
After the pulses interference at the BS of each detection port, 
consider that $(\theta_1^d+\theta_2^d+\theta_3^d)/\pi \mod~2 = 0$ (due to symmetry, we can obtain same result when $(\theta_1^d+\theta_2^d+\theta_3^d)/\pi \mod~2 = 1$), before they entering the detector, we have:   
\begin{equation}
\begin{aligned} \nonumber 
\ket{\psi_{111}}_{L_3R_3L_1R_1L_2R_2} \longrightarrow& \frac{1}{\sqrt{2}}(\ket{101010} + \ket{100101} + \ket{011001} + \ket{010110}) \\
&+\frac{1}{2} e^{\mathbf{i}\theta_3^d}
(\ket{20}-\ket{02}) (\ket{10}+\ket{01})\ket{00} +\frac{1}{2} e^{\mathbf{i}\theta_2^d}(\ket{10}+\ket{01})
\ket{00}
(\ket{20}-\ket{02}) \\
&+\frac{1}{2} e^{\mathbf{i}(\theta_2^d+\theta_3^d)}
(\ket{20}-\ket{02})
\ket{00}(\ket{10}-\ket{01})+\frac{1}{2} e^{\mathbf{i}\theta_1^d}
\ket{00}
(\ket{20}-\ket{02})
(\ket{10}+\ket{01}) \\
&+\frac{1}{2} e^{\mathbf{i}(\theta_1^d+\theta_3^d)}
(\ket{10}-\ket{01})
(\ket{20}-\ket{02})
\ket{00}+\frac{1}{2} e^{\mathbf{i}(\theta_1^d+\theta_2^d)}
\ket{00}
(\ket{10}-\ket{01})
(\ket{20}-\ket{02}), 
\end{aligned}
\end{equation}
\begin{equation}
\begin{aligned}  
\ket{\psi_{110}}_{L_3R_3L_1R_1L_2R_2} \longrightarrow&
\frac{1}{2}
(\ket{10}+\ket{01})(\ket{10}+\ket{01})\ket{00}+\frac{1}{2}e^{\mathbf{i}\theta_2^d}
(\ket{10}+\ket{01})
\ket{00}(\ket{10}-\ket{01})\\
&+\frac{1}{\sqrt{2}}e^{\mathbf{i}\theta_1^d}\ket{00}(\ket{20}-\ket{02})\ket{00} +\frac{1}{2}e^{\mathbf{i}(\theta_1^d+\theta_2^d)}\ket{00}(\ket{10}-\ket{01})
(\ket{10}-\ket{01}),\\
\ket{\psi_{101}}_{L_3R_3L_1R_1L_2R_2}  \longrightarrow&
\frac{1}{2}
(\ket{10}+\ket{01})\ket{00}(\ket{10}+\ket{01})
+\frac{1}{\sqrt{2}}e^{\mathbf{i}\theta_3^d}(\ket{20}-\ket{02})\ket{00}\ket{00} \\
&+\frac{1}{2}e^{\mathbf{i}\theta_1^d}\ket{00}
(\ket{10}-\ket{01})(\ket{10}+\ket{01})
+\frac{1}{2}e^{\mathbf{i}(\theta_1^d+\theta_3^d)}(\ket{10}-\ket{01})
(\ket{10}-\ket{01})\ket{00},\\
\ket{\psi_{011}}_{L_3R_3L_1R_1L_2R_2}   \longrightarrow&
\frac{1}{2}\ket{00}
(\ket{10}+\ket{01})(\ket{10}+\ket{01})+\frac{1}{2}e^{\mathbf{i}\theta_3^d}(\ket{10}-\ket{01})(\ket{10}+\ket{01})\ket{00} \\
&+\frac{1}{\sqrt{2}}e^{\mathbf{i}\theta_2^d}\ket{00}\ket{00} (\ket{20}-\ket{02}) +\frac{1}{2}e^{\mathbf{i}(\theta_2^d+\theta_3^d)}(\ket{10}-\ket{01})\ket{00}
(\ket{10}-\ket{01}),\\ 
\end{aligned}
\end{equation}
\begin{equation}
\begin{aligned}\nonumber
\ket{\psi_{100}}_{L_3R_3L_1R_1L_2R_2}   \longrightarrow&
\frac{1}{\sqrt{2}}
(\ket{10}+\ket{01})\ket{00}\ket{00}+\frac{1}{\sqrt{2}}e^{\mathbf{i}\theta_1^d}
\ket{00}(\ket{10}-
\ket{01})\ket{00},
\\
\ket{\psi_{010}}_{L_3R_3L_1R_1L_2R_2}   \longrightarrow&
\frac{1}{\sqrt{2}}\ket{00}
(\ket{10}+\ket{01})\ket{00} +\frac{1}{\sqrt{2}}e^{\mathbf{i}\theta_2^d}\ket{00}
\ket{00}(\ket{10}-\ket{01}),
\\
\ket{\psi_{001}}_{L_3R_3L_1R_1L_2R_2}   \longrightarrow&
\frac{1}{\sqrt{2}}\ket{00}\ket{00} 
(\ket{10}+\ket{01})+\frac{1}{\sqrt{2}}e^{\mathbf{i}\theta_3^d}(\ket{10}-
\ket{01})\ket{00}\ket{00},
\\
\ket{\psi_{000}}_{L_3R_3L_1R_1L_2R_2}   \longrightarrow&
\ket{00}
\ket{00}
\ket{00}.  
\end{aligned}
\end{equation} 
Therefore, the probability of obtaining a coincident click event can be calculated as:
\begin{equation}
\begin{aligned}
y^{R,R,R} =& (1-p_d)^3\left[
\frac{\eta^3p_d}{16} +
\frac{\eta^2(1-\eta)}{4} \left(\frac{1}{4}p_d + \frac{1}{2}p_d^2\right) + \frac{\eta(1-\eta)^2}{2}p_d^2 +(1-\eta)^3p_d^3 \right],
\\
y^{L,L,L} =& (1-p_d)^3\left[
\frac{\eta^3}{16}+
\frac{\eta^3p_d}{16}+
\frac{\eta^2(1-\eta)}{4} \left(\frac{1}{4}p_d + \frac{1}{2}p_d^2\right) + \frac{\eta(1-\eta)^2}{2}p_d^2 +(1-\eta)^3p_d^3
\right].
\end{aligned}
\end{equation}
\end{widetext}
Due to symmetry, we have $y^{R,R,R} = y^{R,L,L} =  y^{L,R,L} = y^{L,L,R} $ and $y^{L,L,L} = y^{R,R,L} =  y^{R,L,R} =  y^{L,R,R}$.

\subsection{\texorpdfstring{$N=4$}{N=4} with infinite decoy states}
For the $N$ = 4 case, according from Eq.~\eqref{eq_keyrate_infdecoy}, the key length formula is $ 
l_{\rm asy} =s_{1111}^z \left[1-H_2(\phi_{1111}^z)\right]-{\rm leak}_{\rm EC}$. 
The number of single-photon components in the $\boldsymbol{Z}$-basis is:
\begin{equation}
\begin{aligned}
s_{1111}^z =& n_{\rm tot}\frac{(p_\mu p_o \mu e^{-\mu}p_{\rm nc})^4}
{q_{\rm tot}^{P_1}q_{\rm tot}^{P_2}q_{\rm tot}^{P_3}q_{\rm tot}^{P_4}}
(2y_{10}y_{10}y_{10}y_{10}  +12y_{11}y_{10}y_{10}y_{00} +
2y_{11}y_{11}y_{00}y_{00}).
\end{aligned}
\end{equation}
When $N=4$, there are $2^4 = 16$ cases of intensity distribution in the $\boldsymbol{Z}$-basis. The number of single-photon components in the $\boldsymbol{X}$-basis is calculated using the relation $
\frac{s_{1111}^z}{s_{1111}^x} = \frac{\mu^4e^{-4\mu}p_{[\mu,\mu,\mu,\mu]}}{(2\nu e^{-2\nu})^4p_{[2\nu,2\nu,2\nu,2\nu]}},
$, 
where $p_{[\mu,\mu,\mu,\mu]}=16(p_{\mu}p_o)^4$, $p_{[2\nu,2\nu,2\nu,2\nu]}=2p_{\nu}^8/M$.  
The four-party GHZ state $\ket{\Phi_0^+}$ is used as the reference state:
\begin{equation}
\begin{aligned}
\ket{\Phi_0^+}=& \frac{1}{\sqrt{2}}(\ket{+z}^{\otimes4}+\ket{-z}^{\otimes4}) \\
=& \frac{1}{\sqrt{2}}(\ket{+x+x+x+x} +\ket{+x+x-x-x} + \ket{+x-x+x-x} + \ket{+x-x-x+x} 
\\
&+ \ket{-x+x+x-x} + \ket{-x+x-x+x} 
+ \ket{-x-x+x+x} + \ket{-x-x-x-x}).
\end{aligned}
\end{equation}
The number of error event of the single-photon components in the $\boldsymbol{X}$-basis is  
\begin{equation}
\begin{aligned}
t_{1111}^x =&
n_{\rm tot}\left(\frac{2}{M}\right)\frac{(2\nu e^{-2\nu}p_\nu^{2})^4}{q_{\rm tot}^{P_1}q_{\rm tot}^{P_2}q_{\rm tot}^{P_3}q_{\rm tot}^{P_4}}
\times \big[(1-e_d)\mathcal{Y}^{\rm err}_{1111}+  e_d   \mathcal{Y}^{\rm cor}_{1111}  \big],
\end{aligned}
\end{equation} 
where $\mathcal{Y}^{\rm err}_{1111} = 
{y^{R,L,L,L}}+
{y^{L,R,L,L}}+
{y^{L,L,R,L}}+
{y^{L,L,L,R}}+
{y^{R,R,R,L}}+
{y^{R,R,L,R}}+
{y^{R,L,R,R}}+ 
{y^{L,R,R,R}}
$,  $\mathcal{Y}^{\rm cor}_{1111}  =  
{y^{L,L,L,L}}+
{y^{L,L,R,R}}+
{y^{L,R,L,R}}+
{y^{L,R,R,L}}+
{y^{R,L,L,R}}+
{y^{R,L,R,L}}+
{y^{R,R,L,L}}+
{y^{R,R,R,R}}
$, $y^{R(L),R(L),R(L),R(L)}$ is the probability of having detector $R_1(L_1)$, $R_2(L_2)$, $R_3(L_3)$, $R_4(L_4)$  click when input single-photon states in the $\boldsymbol{X}$-basis.

Using the similar method as the three-party scenario,  we have
$  y^{R,L,L,L} = [p_4 (\frac{3}{2}p_d+ 
\frac{1}{2}p_d^2)$ +
$p_3 (\frac{1}{2}p_d+ p_d^2  ) +p_2' (\frac{3}{4}p_d^2
+ \frac{1}{2}p_d^3 ) $+ $ p_2^{''}p_d^2 + \frac{1}{2}p_1p_d^3 
+p_o p_d^4](1-p_d)^4$ and 
$y^{L,L,L,L} =  [p_4(\frac{1}{4}
+\frac{3}{2}p_d+ 
\frac{1}{2}p_d^2 )+ p_3 (\frac{1}{2}p_d+ p_d^2 ) + p_2'(\frac{3}{4}p_d^2+ \frac{1}{2}p_d^3) + p_2''p_d^2
+ \frac{1}{2}p_1p_d^3 
+p_o p_d^4](1-p_d)^4$. 
where 
$p_4= \frac{\eta^4}{16}$,
$p_3=\frac{1}{8}\eta^3(1-\eta)$,
$p_2'=\frac{1}{4}\eta^2(1-\eta)^2 $,
$p_2''= \frac{1}{8}\eta^2(1-\eta)^2$,
$p_1=\frac{1}{2}\eta(1-\eta)^3$ and
$p_o = (1-\eta)^4$. 
Because of the symmetric relation, for the other terms, we have $
y^{R,L,L,L}= y^{L,R,L,L}= y^{L,L,R,L}= y^{L,L,L,R}= y^{R,R,R,L}=y^{R,R,L,R}=y^{R,L,R,R}=y^{L,R,R,R}$  and $y^{L,L,L,L} = y^{L,L,R,R} =  y^{L,R,L,R} =  y^{L,R,R,L}=  y^{R,L,L,R}=  y^{R,L,R,L}=  y^{R,R,L,L}=  y^{R,R,R,R}. $

\subsection{\texorpdfstring{$N=3$}{N=3} 
with three decoy states}  
Here we present the decoy  state calculation formulas for the AMDI-QCKA  with three users and three decoy states. We consider the finite case. The secret key length $l$ against coherent attacks in the finite-size regime can be given as
\begin{equation}
\begin{aligned}
l \ge& ~ \underline{s}_{0}^z + \underline{s}_{111}^z[1-H_2(\overline{\phi}_{111}^z)] -{\rm leak}_{\rm EC}  -\log_2 \frac{4}{\varepsilon_{\rm cor}}  -2\log_2\frac{2}{{\varepsilon'\hat{\varepsilon}}}- 2\log_2\frac{1}{2{\varepsilon_{\rm PA}}},
\end{aligned}
\end{equation}
where we use $\underline{x}$ ($\overline{x}$) to denote the observed upper (lower) bound of $x$. We briefly introduce the formulas of calculating the finite size effect. 
{The Chernoff bound and its variant are utilized to calculate the observed value given the expected value, and the expected value given the observed value, respectively.~\cite{chernoff1952measure, yin2020tight}}. 
Given the expected value $x^*$ and failure probability $\epsilon$, the upper bound and lower bound of the observed value is $\overline{x}=x^{*}+\frac{\beta}{2}+\sqrt{2\beta x^{*}+\frac{\beta^{2}}{4}}$, and $\underline{x}=x^{*}-\sqrt{2\beta x^{*}}$, where $\beta=\ln{\epsilon^{-1}}$. Using the variant of Chernoff bound~\cite{yin2020tight}, we have the upper bound and lower bound of the expected value $x^*$ for a given observed value $x$ and failure probablity $\epsilon$, which can be written as $\overline{x}^{*}=x+\beta+\sqrt{2\beta x+\beta^{2}}$, and
$\underline{x}^{*}=\max\left\{x-\frac{\beta}{2}-\sqrt{2\beta x+\frac{\beta^{2}}{4}},~0\right\}$. 
To obtain the upper bound of the phase error rate in the $\boldsymbol{Z}$-basis, $\overline{\phi}_{111}^z$, the random sampling theorem is utilized~\cite{yin2020tight}, which is given by $	\overline{\chi}\leq\lambda	+\gamma^{U} (n,k,\lambda,\epsilon)$, where
$\gamma^{U}(n,k,\lambda,\epsilon)=\left[\frac{(1-2\lambda)AG}{n+k}+
\sqrt{\frac{A^2G^2}{(n+k)^2}+4\lambda(1-\lambda)G}\right] \left(2+2\frac{A^2G}{(n+k)^2}\right)^{-1}$, with $A=\max\{n,k\}$ and $G=\frac{n+k}{nk}\ln{\frac{n+k}{2\pi nk\lambda(1-\lambda)\epsilon^{2}}}$. 

By using the derivation method in Ref.~\cite{fu2015long}, we have the lower bound of the number of single-photon components:
\begin{widetext}  
\begin{equation}
\begin{aligned}
\underline{s}_{111}^{z*}\ge &\frac{e^{-3\mu}p_{[\mu,\mu,\mu]}}{\nu^3( \mu-\nu)} 
\left[
\mu^4
\left( 
e^{3\nu}\frac{\underline{n}^*_{[\nu,\nu,\nu]}}{p_{[\nu,\nu,\nu]}}
-e^{2\nu}\frac{\overline{n}^*_{[\nu,\nu,o]}}{p_{[\nu,\nu,o]}} - e^{2\nu}\frac{\overline{n}^*_{[\nu,o,\nu]}}{p_{[\nu,o,\nu]}} -  
e^{2\nu}\frac{\overline{n}^*_{[o,\nu,\nu]}}{p_{[o,\nu,\nu]}} +
e^{\nu}\frac{\underline{n}^*_{[\nu,o,o]}}{p_{[\nu,o,o]}} +
e^{\nu}\frac{\underline{n}^*_{[o,\nu,o]}}{p_{[o,\nu,o]}} 
\right.\right.\\
&\left.\left.+e^{\nu}\frac{\underline{n}^*_{[o,o,\nu]}}{p_{[o,o,\nu]}} -
\frac{\overline{n}^*_{[o,o,o]}}{p_{[o,o,o]}}
\right)
-\nu^4
\left( 
e^{3\mu}\frac{\overline{n}^*_{[\mu,\mu,\mu]}}{p_{[\mu,\mu,\mu]}} -  e^{2\mu}\frac{\underline{n}^*_{[\mu,\mu,o]}}{p_{[\mu,\mu,o]}} - e^{2\mu}\frac{\underline{n}^*_{[\mu,o,\mu]}}{p_{[\mu,o,\mu]}} 
- e^{2\mu}\frac{\underline{n}^*_{[o,\mu,\mu]}}{p_{[o,\mu,\mu]}} 
\right.\right.\\
&\left.\left.+
e^{\mu}\frac{\overline{n}^*_{[\mu,o,o]}}{p_{[\mu,o,o]}} +
e^{\mu}\frac{\overline{n}^*_{[o,\mu,o]}}{p_{[o,\mu,o]}} +
e^{\mu}\frac{\overline{n}^*_{[o,o,\mu]}}{p_{[o,o,\mu]}} -
\frac{\underline{n}^*_{[o,o,o]}}{p_{[o,o,o]}}
\right)
\right].
\end{aligned}
\end{equation}
\end{widetext}  
For the data in the $\boldsymbol{Z}$-basis, when the click filtering is not used, we can also use $[\mu,\mu,\nu]$, $[\nu,\mu,\mu]$, $[\mu,\nu,\mu]$, $[\mu,\nu,\nu]$, $[\nu,\mu,\nu]$, $[\nu,\nu,\mu]$, $[\nu,\nu,\nu]$ events to generate secret keys. In this case, we can replace term $e^{-3\mu} p_{[\mu,\mu,\mu]}/\nu^3$ with $ (\mu^3e^{-3\mu} p_{[\mu,\mu,\mu]}+3\mu^2\nu e^{-2\mu-\nu} p_{[\mu,\mu,\nu]}+3\mu\nu^2 e^{-\mu-2\nu} p_{[\mu,\nu,\nu]}+\nu^3 e^{-3\nu} p_{[\mu,\mu,\mu]})/(\mu^3\nu^3)$. 

The number of vacuum components, which is defined as one user sends vacuum state, can be given by 
\begin{equation}
\begin{aligned}
\textcolor{red}{
\underline{s}_{0}^{z*}\geq}& e^{-\mu } p_{[\mu,\mu,\mu]}
\frac{\underline{n}_{[o,\mu, \mu]}^{ *} }{p_{[o,\mu,\mu]}}.
\end{aligned}
\end{equation} 
and $\underline{s}_{111}^{x*}$ can be calculated using the relation $ 
\frac{s_{111}^{z*}}{s_{111}^{x*}} = \frac{\mu^3e^{-3\mu}p_{[\mu,\mu,\mu]}}{(2\nu e^{-2\nu})^3p_{[2\nu,2\nu,2\nu]}}$, 
the upper bound of single-photon state errors of the $\boldsymbol{X}$-basis is  
\begin{equation}
\begin{aligned}\label{mo}
\overline{t}_{111}^{x*} \le & e^{-6\nu}p_{[2\nu,2\nu,2\nu]}
\bigg(e^{6\nu}\frac{\overline{m}_{[2 \nu,2 \nu,2 \nu]}^{*}}{p_{[2 \nu,2 \nu,2 \nu]}}-e^{4\nu}\frac{\underline{n}_{[2 \nu,2 \nu,o]}^{*}}{2p_{[2 \nu,2 \nu,o]}}
-e^{4\nu}\frac{\underline{n}_{[2 \nu,o,2 \nu]}^{*}}{2p_{[2 \nu,o,2 \nu]}}-
e^{4\nu}\frac{\underline{n}_{[o,2 \nu,2 \nu]}^{*}}{2p_{[o,2 \nu,2 \nu]}} 
+e^{2\nu}\frac{\overline{n}_{[2 \nu,o,o]}^{*}}{2p_{[2 \nu,o,o]}} \\
&+e^{2\nu}\frac{\overline{n}_{[o,2 \nu,o]}^{*}}{2p_{[o,2 \nu,o]}}
+e^{2\nu}\frac{\overline{n}_{[o,o,2 \nu]}^{*}}{2p_{[o,o,2 \nu]}}
-\frac{\underline{n}_{[o,o,o]}^{*}}{2p_{[o,o,o]}}
\bigg) ,
\end{aligned}
\end{equation}
where $m_{[k_1^{\rm tot},k_2^{\rm tot},k_3^{\rm tot}]}$ denotes the number of errors  of $[k_1^{\rm tot},k_2^{\rm tot},k_3^{\rm tot}]$ event. Using the random sampling without replacement theorem, with a failure probability $\varepsilon_e$, the upper bound of the  phase error rate of single-photon components in the $\boldsymbol{Z}$-basis is 
\begin{equation}
\begin{aligned}    \overline{\phi}_{111}^{z}\leq& \overline{e}_{111}^x	+\gamma \left(\underline{s}_{111}^z,\underline{s}_{111}^x,\overline{e}_{111}^x,\varepsilon_e\right),
    \label{phi_11z_stop}
\end{aligned} 
\end{equation}
where $\overline{e}_{111}^x = \overline{t}_{111}^x/{\underline{s}_{111}^x}$ is the upper bound of bit error rate in $\boldsymbol{X}$-basis.
 
\section{Simulation formulas}
\label{sec_app_simu}
In this section, we provide the simulation formulas for the AMDI-QCKA protocol with $N$ users.  We consider the symmetric channel. Each user $U_i$ sends phase-randomized weak coherent pulse with random intensity $k_i\in \{\mu, \nu, 0\}$. we define the subscript ${N+1}:=1$ for cyclic structure of notation.
When the coherent state from $U_i$ ($i\in{1,2,...,N}$) arrive at the center relay and pass through the $1\times 2$ BS, the joint quantum states can be written as
\begin{widetext}
\begin{equation}
\begin{aligned}
\bigotimes_{i=1}^N \ket{e^{\mathbf{i}\theta_i}\sqrt{k_i\eta}} 
\xrightarrow[]{1\times 2~\rm BS} 
&  
\bigotimes_{i=1}^{N} 
\left(\ket{e^{\mathbf{i}\theta_{i}}\sqrt{\frac{k_i\eta}{2}}}_{P_{i}} 
\ket{e^{\mathbf{i}\theta_{i+1}}\sqrt{\frac{k_{i+1}\eta}{2}}}_{P_{i}} \right)  \\
\xrightarrow[]{2\times 2~\rm BS} 
& \bigotimes_{i=1}^{N} 
\left(\ket{e^{\mathbf{i}\theta_{i}}\frac{\sqrt{k_i\eta}}{2}+e^{\mathbf{i}\theta_{i+1}}\frac{\sqrt{k_{i+1}\eta}}{2}}_{L_{i}}  
\ket{e^{\mathbf{i}\theta_{i}}\frac{\sqrt{k_i\eta}}{2}-e^{\mathbf{i}\theta_{i+1}}\frac{\sqrt{k_{i+1}\eta}}{2}}_{R_{i}} \right),
\end{aligned}
\end{equation}
\end{widetext}
where $\theta_i$ is the random phase of $U_i$, $\eta$ is the channel transmittance from $U_i$ to the center relay, $P_i$ represent the $i$-th detection port. $L_i$ ($R_i$) represents the left (right) side single-photon detector. Therefore, the probability of the detector $L_i$ and $R_i$ have a click is
$E_{L_i} = 1 - (1 - p_d)e^{-(\alpha_i + \beta_i\cos\hat{\theta}_{i})}$ and
$E_{R_i} = 1 - (1 - p_d)e^{-(\alpha_i - \beta_i\cos\hat{\theta}_{i})}$, respectively, 
where $\alpha_i = \frac{\eta(k_i+k_{i+1})}{4}$, $\beta_i = \frac{\eta\sqrt{k_ik_{i+1}}}{2}$, $\hat{\theta}_{i} =\theta_{i}-\theta_{i+1}$. 
Then, for each detector we have the probability of having a single click event 
\begin{equation}
\begin{aligned}
q^{\hat{\theta}_i,L_i}_{\vec{k}} =& E_{L_i}(1-E_{R_i})\prod_{j=1, j\neq i}^N
(1-E_{R_j})(1-E_{L_j}) \\
=& (1-p_d)^{2N-1}
\exp{\left[-\left(\sum\limits_{j=1}^N2\alpha_j
-\alpha_i-\beta_i \cos\hat{\theta}_{i}\right)\right]} -(1-p_d)^{2N}\exp{\left(-\sum\limits_{j=1}^N2\alpha_j\right)}, 
\\
q^{\hat{\theta}_i,R_i}_{\vec{k}} =& E_{R_i}(1-E_{L_i})\prod_{j=1, j\neq i}^N
(1-E_{R_j})(1-E_{L_j}) \\
=& (1-p_d)^{2N-1}
\exp{\left[-\left(\sum\limits_{j=1}^N2\alpha_j
-\alpha_i+\beta_i \cos\hat{\theta}_{i}\right)\right]} -(1-p_d)^{2N}\exp{\left(-\sum\limits_{j=1}^N2\alpha_j\right)}. 
\end{aligned}
\end{equation}  
Here, $q^{\hat{\theta}_j,L_i(R_i)}$ is the function of $\hat{\theta_j}$. 
Therefore, given intensity $\vec{k}$, the overall probability of having a single click is 
\begin{equation}
\begin{aligned} \label{eq_qveck}
q_{\vec{k}}
=& \sum_{i=1}^{N} 
\frac{1}{2\pi} \int_0^{2\pi}
\left(
q^{\hat{\theta}_i,L_i} + 
q^{\hat{\theta}_i,R_i} \right)
d\hat{\theta}_i 
\\
=& 2(1-p_d)^{2N-1}\sum_{i=1}^{N} 
I_0(\beta_i)\exp{\left(-(\sum\limits_{j=1}^N2\alpha_j
-\alpha_i)\right)} - 2N(1-p_d)^{2N}\exp{\left(-\sum\limits_{j=1}^N2\alpha_j\right)},
\end{aligned}
\end{equation}   
where $\vec{k}=\{k_1,k_2,...k_N\}$ represents the intensity of all users.  To calculate the integration of $d\hat{\theta}_i$, we use the formulas for zero-order modified Bessel function of the first kind $I_0(\sqrt{a^2 +b^2} x) = \frac{1}{2\pi} \int_{0}^{2\pi} e^{-x (a\cos\theta+b\sin\theta)} d\theta$. 
Separately consider the each detection port, we also have the probability that only detection port $P_i$ clicks: $q^{P_i}_{\vec{k}} = 2(1-p_d)^{2N-1} 
I_0(\beta_i)e^{-(\sum\limits_{j=1}^N2\alpha_j
-\alpha_i)} - 2(1-p_d)^{2N}e^{-\sum\limits_{j=1}^N2\alpha_j}$.  The above probabilities are a function of the light intensity $k_i$. By taking all the possible intensity combinations into account, we have the total probability of having a click at 
a specific detection port, which can be written as $q_{\rm tot}^{P_i} =\sum\limits_{\vec{k}}   \vec{p}_k q^{P_i}_{\vec{k}}$, where $\vec{p}_k = \prod\limits_{i=1}^N p_{k_i}$  is the set of the probability for the users   choosing intensity $\{k_1,k_2,...k_N\}$. In addition, the total probability of having a click event is  $q_{\rm tot} =\sum\limits_{\vec{k}} \vec{p}_k q_{\vec{k}} 
$. 

\begin{figure}[t]
\includegraphics[width=0.48\textwidth]{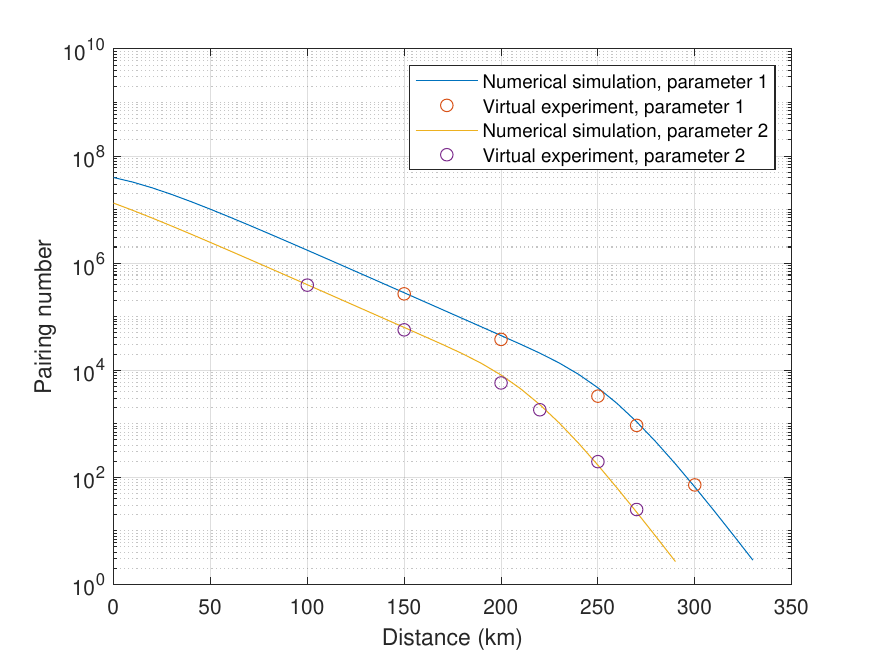}
\caption{Simulation result of the number of pairing event, where the circles represent $n_{\rm tot}$ obtained in virtual experiments, and the data points represent $n_{\rm tot}$ numerically calculated. We considered two set of parameters with different intensity.
The comparison between our numerical formula and the virtual experiment results demonstrates a high level of agreement, indicating the accuracy of the numerical simulations.  
}\label{fig_supp_pairnum}
\end{figure}

So far, we have finished the detection probability calculation of a single time bin. In the pairing step, detection events that occur in $N$ different time bins at different detector ports are paired. Then we calculate the number of pairing event, $n_{\rm tot}$. When the global phase of each user are locked, almost all the detection events can be paired. {
The number of pairing events is limited by the port with the fewest clicks.}
Thus, we have $n_{\rm tot}= \mathcal{N}
\min q_{\rm tot}^{P_i} $. 
{
In symmetric case, by using $ q_{\rm tot}^{P_1} = q_{\rm tot}^{P_2} = \dots = q_{\rm tot}^{P_N} =q_{\rm tot}/N$, 
$n_{\rm tot}$ can also be expressed as $n_{\rm tot}=\mathcal{N} q_{\rm tot}/N$.
}
When the global phase locking is not utilized, the maximum pairing time interval between the earliest and the latest should be less than $T_c$ to reduce the 
$\boldsymbol{X}$-basis error rate. Given the earliest time bin with a click event at $P_i$, the probability of finding at least $N-1$ click events in the other $N-1$ different ports is $
\prod\limits_{j=1, j\neq i}^N
[1-(1-q_{\rm tot}^{P_j})^{\mathcal{N}_{T_c}}]
$, where $\mathcal{N}_{T_c}$ is the number of pulses that each user send within $T_c$. The average number of time bins that are required to obtain a pairing event is $1 + (N-1)/\prod\limits_{j=1, j\neq i}^N
[1-(1-q_{\rm tot}^{P_j})^{\mathcal{N}_{T_c}}]$.
Then, by taking into account all possible detection ports, we have the number pairing event 
\begin{equation} \label{eq_ppair} 
n_{\rm tot}=
\sum\limits_{i=1}^N  \mathcal{N}
q_{\rm tot}^{P_i} \left\{1 + \frac{N-1}{\prod\limits_{j=1,j\neq i}^N
\left[1-(1-q_{\rm tot}^{P_j})^{\mathcal{N}_{T_c}}\right]}\right\}^{-1}.
\end{equation}
To show the accuracy of Eq.~\eqref{eq_ppair}, we perform a virtual experiment using computer, recorded $n_{\rm tot}$ and compared with that computed using the above formula. As shown in Fig.~\ref{fig_supp_pairnum}, the result of Eq.~\eqref{eq_ppair} fit well with the virtual experiment. 
In simulation, we assume the mean pairing time interval (the maximum time interval between the first and the last time bin) is $T_{\rm mean} = T_c/2$. 
 
Then, we consider the number of the paired event with total intensities 
$\vec{K} = [k_1^{\rm tot}, k_2^{\rm tot}, ...,k_N^{\rm tot}]$ (except for $\vec{K}_x = [2 \nu,2 \nu,..., 2 \nu]$): 
\begin{equation}
\begin{aligned}\label{eq_nk1k2k3}
n_{\vec{K}} = & n_{\rm tot}
\sum\limits_{\substack{k_{i}^{e}+k_{i}^{l}= k_{i}^{\rm tot}\\
\forall i\in \{1,...,N\}}}
\prod\limits_{i=1}^{N}   
\left(     \frac{p_{k_i^l}p_{k_{i+1}^e}q_{k_i^lk_{i+1}^e}^{P_i}}{q_{\rm tot}^{P_i}}
\right),
\end{aligned} 
\end{equation}
where  
$q_{k_i^lk_{i+1}^e}^{P_i}  =\sum\limits_{\substack{k_j \\ j \notin \{i, i+1\}}} 
p_{\vec{k}_j}q_{\vec{k}}^{P_i}$, $\vec{k}_j$ is the intensity set excluding $k_i^l$ and $k_{i+1}^e$. For example, in $N = 3$ and $N =4$  cases, for detection port $P_1$, we have 
$q_{k_1^lk_2^e}^{P_1} = \sum\limits_{k_3}p_{k_3} q^{P_1}_{k_1^l k_2^e k_3}$ and  $q_{k_1^lk_2^e}^{P_1} = \sum\limits_{k_3,k_4}p_{k_3}p_{k_4} q^{P_1}_{k_1^lk_2^ek_3k_4}$, respectively. Note that in calculating, the click event of each port is mainly relevant to the intensity and phase of two neighbor users. For the given detection port $P_1$, the pairing event $n_{[2 \nu, 2 \nu, 2 \nu]}$ requires $U_1$ and $U_2$ both send $\nu$ at the time bin when $P_1$ clicks, while $U_3$ may choose any intensities including $\mu$, $\nu$ and $0$. The probability of having a click at $P_1$ is the function of intensities $k_1$, $k_2$, $k_3$. In calculation, we should first sum all $U_3$'s possible intensities.  
 
The $\boldsymbol{Z}$-basis marginal bit error rate between $U_1$ and $U_i$ can be written as $E_{1,i}^z=m_{1,i}^z /n_{\vec{K_z}} $, where $\vec{K}_z = [\mu,\mu,...,\mu]$ is the intensity set of $\boldsymbol{Z}$-basis, and
the number of  $\boldsymbol{Z}$-basis error event
$m^z_{1,i}$ can be given as 
\begin{equation}
\begin{aligned}
m^z_{1,i} =&~n_{\rm tot} \frac{(p_{\mu}p_o)^N}{\prod\limits_{j=1}^N q_{\rm tot}^{P_j}}
\sum\limits_{ (k_1^e,k_1^l, k_i^e,k_i^l)\in  \mathcal{K} }
\prod_{j=1}^N q_{k_j^lk_{j+1}^e}^{P_j}
\end{aligned}
\end{equation} 
where $\mathcal{K} = \{ (k_1^e, k_1^l, k_i^e, k_i^l) \mid (k_1^e, k_1^l), (k_i^e, k_i^l) \in \{(o, \mu), (\mu, o)\}, \ (k_1^e, k_1^l) \neq (k_i^e, k_i^l) \}$,  and we take $U_1$ as the reference. In simulation, we assume the marginal bit error between each users are equal in the symmetric case. 

The number of $\boldsymbol{X}$-basis event, $n_{\vec{K}_x}$, 
where $\vec{K}_x = [2\nu, 2\nu,...,2\nu]$, can be given as 
\begin{equation}
\begin{aligned}
n_{\vec{K}_x} = 
\frac{2}{M}
\frac{n_{\rm tot}p_{\vec{K}_x}}{(2\pi)^N}
\int_0^{2\pi}  \left(
\frac{q_{\nu\nu}^{-(\sum\limits_{i=2}^N \hat{\theta}_i), P_1}}{q_{\rm tot}^{P_1}}
\prod_{i=2}^N \frac{q_{\nu\nu}^{\hat{\theta}_i,P_i}}{q_{\rm tot}^{P_i}}
\right)
 d\vec{\hat{\theta}}_i .
\end{aligned} 
\end{equation}
where $\vec{\hat{\theta}}_i = \{\hat{\theta}_2, \hat{\theta}_3, ..., \hat{\theta}_N\}$, $q_{\nu\nu}^{\hat{\theta}_{i},P_i} =\sum\limits_{\substack{k_j \\ j \notin \{i, i+1\}}}  p_{\vec{k}} (q^{\hat{\theta}_{i},L_i}_{\vec{k}}+ q^{\hat{\theta}_{i},R_i}_{\vec{k}})$.
The global phase of each user arriving at Eve is continuous and random, resulting in the integration from $0$ to $2\pi$. The total number of errors in the $\boldsymbol{X}$-basis can be written as       
\begin{equation}
\begin{aligned} \label{eq_app_mx}
m_{\vec{K}_x}\!\!\!=&
\frac{2}{M}
\frac{n_{\rm tot}p_{\vec{K}_x}}{(2\pi)^N\prod_{i=1}^N 
q_{\rm tot}^{P_i}}
\!\int_0^{2\pi} \!\!\!
\left[
(1-e_d)\mathcal{Y}_{\vec{K}_x}^{ \rm err} \!\! +  e_d \mathcal{Y}_{\vec{K}_x}^{\rm cor} \right] d\vec{\hat{\theta}}_i,
\end{aligned} 
\end{equation} 
where $\mathcal{Y}_{\vec{K}_x}^{ \rm err}$ ($\mathcal{Y}_{\vec{K}_x}^{ \rm cor}$) denotes the probability of obtaining an error (correct) coincidence
click event given intensity $[2\nu,2\nu,...,2\nu]$ as input, which can be given as $
\mathcal{Y}_{\vec{K}_x}^{ \rm err} = \sum\limits_{S^-} \prod\limits_{i=1}^N y_{\nu\nu}^{\theta_i,r_i} $
and   $
\mathcal{Y}_{\vec{K}_x}^{ \rm cor} = \sum\limits_{S^+}  \prod\limits_{i=1}^N y_{\nu\nu}^{\theta_i,r_i} $,
with $\theta_1 = \delta-(\sum\limits_{i=2}^N \hat{\theta}_i)$, $\delta=T_{\rm mean}(2\pi \Delta f + \omega_{\rm fiber})$ is the phase misalignment, $\Delta f$ is the laser frequency difference, $\omega_{\rm fiber}$ is fiber phase drift rate. $e_d$ is the $\boldsymbol{X}$-basis misalignment error rate.  $\mathcal{S^+}: r_1 \oplus r_2 \oplus ... \oplus r_N =
(N-1~\rm {mod}~2)$ and $\mathcal{S^-}: r_1 \oplus r_2 \oplus ... \oplus r_N =
(N~\rm {mod}~2)$, $r_i \in \{0,1\}$ corresponding to detector $\{L_i,R_i\}$ clicks, respectively. 

The intrinsic interference error rate for three-party is approximately $37.5\%$. Let $s_{a_1a_2a_3}$ represents there are $a_i$ photons from $U_i$. The bit error rate in the $\boldsymbol{X}$-basis can be written as $E_X\approx\frac{6e_0s_{210}}{s_{111}+6s_{210}}\approx \frac{3}{8} =37.5\%$, where $e_0=0.5$ is the  vacuum state error rate. Similarly, the intrinsic interference error rate for four-party case in the $\boldsymbol{X}$-basis can be approximately written as $E_X\approx\frac{12e_0s_{2110}} {s_{1111}+12s_{2110}}\approx \frac{3}{7} 
\approx42.86\%$.

In our protocol, when data click filtering method is utilized, for numerical simulations, we have  
\begin{equation}
\begin{aligned}
p_{[k_{1}^{tot},k_{2}^{tot},...,k_{N}^{tot}]}= \frac{1}{p_s^3}
\sum\limits_{\substack{k_{i}^{e}+k_{i}^{l}= k_{i}^{\rm tot}\\
\forall i\in \{1,...,N\}}}\prod\limits_{i=1}^N  p_{k_{i}^{e}}p_{k_{i}^{l}},
\end{aligned}
\end{equation} 
where \textcolor{red}{
$p_s=1-2p_\mu p_\nu$}.

\section{Protocol scaling}
\label{sec_app_scaling}
In this section, we present a qualitative analysis for the key rate scaling of the AMDI-QCKA. We consider the symmetric channels in asymptotic limit with infinite data
size. 
In this case, the detection  probabilities at all detection ports are equal: $ q_{\rm tot}^{P_1} = q_{\rm tot}^{P_2} = \dots = q_{\rm tot}^{P_N} =q_{\rm tot}/N$.
According from Eq.~\eqref{eq_qveck}, the overall probability of having a single click is 
\begin{equation}
\begin{aligned} 
q_{\rm tot}
=~& 2NI_0\left(\frac{\mu_{\rm mean}\eta}{2}\right)(1-p_d)^{2N-1}e^{-\frac{(2N-1)\mu_{\rm mean}\eta}{2}} - 2N(1-p_d)^{2N}e^{-N\mu_{\rm mean}\eta} \\
\approx~& 
2N\left[1+\frac{1}{4}\left(\frac{\mu_{\rm mean}\eta}{2}\right)^2\right]e^{-\frac{(2N-1)\mu_{\rm mean}\eta}{2}}  - 2N e^{-N\mu_{\rm mean}\eta} \\
\approx~& N\mu_{\rm mean}\eta\left(1+\frac{\mu_{\rm mean}\eta}{8}\right) \\
\approx~& N\mu_{\rm mean}\eta, 
\end{aligned}
\end{equation}  
where $\mu_{\rm{mean}}$ is the mean intensity of each user, and we ignored the second-order term and dark count rate $p_d$.
With Eq.~\eqref{eq_ppair}, when the click rate is high such that
one can always obtain at least one click event for each detection port within the coherence time $T_c$, i.e., $q_{\rm tot}^{P_i}\mathcal{N}_{T_c}\ge 1$, the number of pairing event is 
\begin{equation} 
n_{\rm tot}\approx
\sum\limits_{i=1}^N 
\frac{\mathcal{N}
q_{\rm tot}^{P_i} }{N}
\approx \mathcal{N}\mu_{\rm mean} \eta.
\end{equation}
This shows that the number of paired event scales linearly with $\eta$. As a result, in asymptotic regime,  the key rate exhibits an   $O(\eta)$ scaling.

\section{Mermin value} \label{sec_app_mermin}
According from local realism, the Mermin's inequality is~\cite{mermin1990simple,fu2015long}
\begin{equation}
\begin{aligned} 
M = \langle XXX \rangle - \langle XYY \rangle- \langle YXY \rangle- \langle YYX \rangle \le 2,
\end{aligned}
\end{equation}
where $X$ and $Y$ are the Pauli $\sigma_x$ and $\sigma_y$ operators, $M$ is the Mermin value.  For the tripartite GHZ states $\ket{\Phi_0^+}=\frac{1}{\sqrt{2}}(\ket{+z+z+z}+\ket{-z-z-z})$ measured in ideal case, the maximum Mermin value is 4. By using the decoy state method to estimate the single-photon components, we have
\begin{equation}
\begin{aligned} 
M_{3}^{\Phi_0^+} \!\!=\!\! \langle XXX \rangle_{111}^{\Phi_0^+}\!\!-\langle XYY \rangle_{111}^{\Phi_0^+}\!\!- \langle YXY \rangle_{111}^{\Phi_0^+}\!\!- \langle YYX \rangle_{111}^{\Phi_0^+},
\end{aligned}
\end{equation}
where $ \langle XXX \rangle_{111}^{\Phi_0^+}$ is the expected value of the single-photon component, when all the users send quantum state in the $\boldsymbol{X}$-basis and cause a successful ${\Phi_0^+}$ measurement result. The expected value of $ \langle XXX \rangle_{111}^{\Phi_0^+}$ can be given as
\begin{equation}
\begin{aligned} 
\langle XXX \rangle_{111}^{\Phi_0^+} = 
\frac{s_{\rm cor}^{\Phi_0^+} - s_{\rm err}^{\Phi_0^+}}{s_{\rm cor}^{\Phi_0^+} + s_{\rm err}^{\Phi_0^+}},
\end{aligned}
\end{equation}
where $s_{\rm cor}^{\Phi_0^+} + s_{\rm err}^{\Phi_0^+}$ is the total number of single-photon component projected to $\ket{\Phi_0^+}$ in the $\boldsymbol{X}$-basis, $s_{\rm cor}^{\Phi_0^+} = s_{+++}^{\Phi_0^+}+s_{+--}^{\Phi_0^+}+s_{-+-}^{\Phi_0^+}+s_{--+}^{\Phi_0^+}$, $s_{\rm err}^{\Phi_0^+} = s_{++-}^{\Phi_0^+}+s_{+-+}^{\Phi_0^+}+s_{-++}^{\Phi_0^+}+s_{---}^{\Phi_0^+}$, and the subscript
$+(-)$ denotes the user sends a $\ket{+}(\ket{-})$ state. It can be prove that $\langle XXX \rangle_{111}^{\Phi_0^+}$ = $-\langle XYY \rangle_{111}^{\Phi_0^+}$ = $-\langle YXY \rangle_{111}^{\Phi_0^+}$ = $-\langle YYX \rangle_{111}^{\Phi_0^+}$. Due to symmetry, we also have $s_{+++}^{\Phi_0^+} = s_{+--}^{\Phi_0^+} = s_{-+-}^{\Phi_0^+} = s_{--+}^{\Phi_0^+}$ and $s_{++-}^{\Phi_0^+} = s_{+-+}^{\Phi_0^+} = s_{-++}^{\Phi_0^+} = s_{---}^{\Phi_0^+}$.
Therefore, in symmetric case, the lower bound of the Mermin value can be given as
\begin{equation}
\begin{aligned} 
\underline{M}_{111}^{\Phi_0^+} = 4\langle XXX\rangle _{111}^{\Phi_0^+L}=4 \times \frac{\underline{s}_{+++}^{\Phi_0^+}-\overline{s}_{---}^{\Phi_0^+}}{\overline{s}_{+++}^{\Phi_0^+}+\overline{s}_{---}^{\Phi_0^+}}.
\end{aligned}
\end{equation}
By using the decoy state method, we have 
\begin{widetext}  
\begin{equation}
\begin{aligned}
s_{+++}^{\Phi_0^+*}\ge &\frac{e^{-6\nu}p_{[2\nu,2\nu,2\nu]}}{\mu^3(\mu-\nu)} \!\!\!
\left[\!
\mu^4
\left( \!
e^{6\nu}\frac{n^{+++\Phi_0^+}_{[2\nu,2\nu,2\nu]}}{p_{[2\nu,2\nu,2\nu]}}
\!-\!e^{4\nu}\frac{n^{+++\Phi_0^+}_{[2\nu,2\nu,o]}}{p_{[2\nu,2\nu,o]}} \!- \!e^{4\nu}\frac{n^{+++\Phi_0^+}_{[2\nu,o,2\nu]}}{p_{[2\nu,o,2\nu]}} \!- \! 
e^{4\nu}\frac{n^{+++\Phi_0^+}_{[o,2\nu,2\nu]}}{p_{[o,2\nu,2\nu]}} \!+\!
e^{2\nu}\frac{n^{+++\Phi_0^+}_{[2\nu,o,o]}}{p_{[2\nu,o,o]}} \!+\!
e^{2\nu}\frac{n^{+++\Phi_0^+}_{[o,2\nu,o]}}{p_{[o,2\nu,o]}} 
\right.\right.\\
&\left.\left.+e^{2\nu}\frac{n^{\Phi_0^+}_{[o,o,2\nu]}}{p_{[o,o,2\nu]}} -
\frac{n^{+++\Phi_0^+}_{[o,o,o]}}{p_{[o,o,o]}}
\right)
-\nu^4
\left( 
e^{6\mu}\frac{n^{+++\Phi_0^+}_{[2\mu,2\mu,2\mu]}}{p_{[2\mu,2\mu,2\mu]}} -  e^{4\mu}\frac{n^{+++\Phi_0^+}_{[2\mu,2\mu,o]}}{p_{[2\mu,2\mu,o]}} - e^{4\mu}\frac{n^{+++\Phi_0^+}_{[2\mu,o,2\mu]}}{p_{[2\mu,o,2\mu]}} 
- e^{4\mu}\frac{n^{+++\Phi_0^+}_{[o,2\mu,2\mu]}}{p_{[o,2\mu,2\mu]}} 
\right.\right.\\
&\left.\left.+
e^{2\mu}\frac{n^{+++\Phi_0^+}_{[2\mu,o,o]}}{p_{[2\mu,o,o]}} +
e^{2\mu}\frac{n^{+++\Phi_0^+}_{[o,2\mu,o]}}{p_{[o,2\mu,o]}} +
e^{2\mu}\frac{n^{+++\Phi_0^+}_{[o,o,2\mu]}}{p_{[o,o,2\mu]}} -
\frac{n^{+++\Phi_0^+}_{[o,o,o]}}{p_{[o,o,o]}}
\right) 
\right],
\end{aligned}
\end{equation} 
\begin{equation}
\begin{aligned}
s_{+++}^{\Phi_0^+*}\le &{e^{-6\nu} p_{[2\nu,2\nu,2\nu]}
} 
\left[ 
e^{6\nu}\frac{n^{+++\Phi_0^+}_{[2\nu,2\nu,2\nu]}}{p_{[2\nu,2\nu,2\nu]}}
-
e^{4\nu}\left(\frac{n^{+++\Phi_0^+}_{[2\nu,2\nu,o]}}{p_{[2\nu,2\nu,o]}} +\frac{n^{+++\Phi_0^+}_{[2\nu,o,2\nu]}}{p_{[2\nu,o,2\nu]}} + \frac{n^{+++\Phi_0^+}_{[o,2\nu,2\nu]}}{p_{[o,2\nu,2\nu]}}\right) \right. \\
& \left.+
e^{2\nu}\left(\frac{n^{+++\Phi_0^+}_{[2\nu,o,o]}}{p_{[2\nu,o,o]}} +
\frac{n^{+++\Phi_0^+}_{[o,2\nu,o]}}{p_{[o,2\nu,o]}} 
+\frac{n^{+++\Phi_0^+}_{[o,o,2\nu]}}{p_{[o,o,2\nu]}}\right) -
\frac{n^{+++\Phi_0^+}_{[o,o,o]}}{p_{[o,o,o]}}
\right] ,
\end{aligned}
\end{equation}
\begin{equation}
\begin{aligned}
s_{---}^{\Phi_0^+}\le &
e^{-6\nu} p_{[2\nu,2\nu,2\nu]} 
\left[
e^{6\nu}\frac{n^{---\Phi_0^+}_{[2\nu,2\nu,2\nu]}}{p_{[2\nu,2\nu,2\nu]}}
-e^{4\nu}
\left(\frac{n^{---\Phi_0^+}_{[2\nu,2\nu,o]}}{p_{[2\nu,2\nu,o]}}+ \frac{n^{---\Phi_0^+}_{[2\nu,o,2\nu]}}{p_{[2\nu,o,2\nu]}} +  
\frac{n^{---\Phi_0^+}_{[o,2\nu,2\nu]}}{p_{[o,2\nu,2\nu]}}\right) \right.\\
& \left. +e^{2\nu}\left(\frac{n^{---\Phi_0^+}_{[2\nu,o,o]}}{p_{[2\nu,o,o]}} +
\frac{n^{---\Phi_0^+}_{[o,2\nu,o]}}{p_{[o,2\nu,o]}} 
+\frac{n^{---\Phi_0^+}_{[o,o,2\nu]}}{p_{[o,o,2\nu]}}\right) -
\frac{n^{---\Phi_0^+}_{[o,o,o]}}{p_{[o,o,o]}}
\right],
\end{aligned}
\end{equation}

\end{widetext}  
where $n^{+++\Phi_0^+}_{[2\nu,2\nu,2\nu]}$ is the number of $[2\nu,2\nu,2\nu]$ event with all the users send $\ket{+x}$ state and cause a $\ket{\Phi_0^+}$ detection event. 
\begin{equation}
\begin{aligned} 
n^{---\Phi_0^+}_{[2\nu,2\nu,2\nu]}=&\frac{n_{\rm tot}p_{ \nu}^6}{16M\pi^2q_{\rm tot}^{P_1}q_{\rm tot}^{P_2}q_{\rm tot}^{P_3}}\int_0^{2\pi} 
\mathcal{Y}_{[2\nu,2\nu,2\nu]}^{ \rm err}
d \vec{\hat{\theta}}, \\
n^{+++\Phi_0^+}_{[2\nu,2\nu,2\nu]}=&\frac{n_{\rm tot}p_{ \nu}^6}{16M\pi^2q_{\rm tot}^{P_1}q_{\rm tot}^{P_2}q_{\rm tot}^{P_3}}\int_0^{2\pi} 
\mathcal{Y}_{[2\nu,2\nu,2\nu]}^{\rm cor}
d \vec{\hat{\theta}},
\end{aligned} 
\end{equation}
where $\mathcal{Y}_{[2\nu,2\nu,2\nu]}^{\rm cor}$ and $\mathcal{Y}_{[2\nu,2\nu,2\nu]}^{\rm err}$ is given in Eq.~\eqref{eq_app_mx}.
For the other terms, we have $n_{[k_1^{\rm tot}, k_2^{\rm tot}, k_3^{\rm tot}]}^{+++\Phi_0^+}$ =$ \frac{1}{16}n_{[k_1^{\rm tot}, k_2^{\rm tot}, k_3^{\rm tot}]}$.


%